\documentclass[10pt,letterpaper]{article}

\usepackage[T1]{fontenc}
\usepackage[utf8]{inputenc}
\usepackage{amsmath}
\usepackage{newtxtext,newtxmath}
\usepackage[scale=0.9]{sourcecodepro}
\usepackage[letterpaper,textwidth=5.5in,textheight=9in,top=1in,headheight=12pt,headsep=25pt,footskip=30pt]{geometry}
\usepackage[expansion=false]{microtype}
\usepackage{booktabs,tabularx,array}
\usepackage{xcolor}
\usepackage{graphicx}
\usepackage{changepage}
\usepackage{needspace}
\usepackage{caption}
\usepackage[numbers,sort&compress]{natbib}
\usepackage{enumitem}
\definecolor{iaInk}{HTML}{0D0C16}
\definecolor{iaDeep}{HTML}{20213F}
\definecolor{iaViolet}{HTML}{4442B1}
\definecolor{iaSlate}{HTML}{3A395A}
\definecolor{iaOrange}{HTML}{E7844D}
\usepackage[colorlinks=true,linkcolor=iaViolet,citecolor=iaViolet,urlcolor=iaViolet]{hyperref}
\hypersetup{pdftitle={Test-Time Communication in Trading Agent Swarms},
  pdfauthor={Avid Fayaz, Harry Zhang, Håvard Alstadheim, Erik Nordlöf, Vilhelm Karlin, André Silva}}

\makeatletter
\renewcommand\normalsize{\@setfontsize\normalsize\@xpt\@xipt\abovedisplayskip 7\p@ \@plus2\p@ \@minus5\p@ \abovedisplayshortskip \z@ \@plus3\p@ \belowdisplayshortskip 4\p@ \@plus3\p@ \@minus3\p@ \belowdisplayskip \abovedisplayskip \let\@listi\@listI}
\normalsize
\renewcommand\small{\@setfontsize\small\@ixpt\@xpt\abovedisplayskip 6\p@ \@plus1.5\p@ \@minus4\p@ \abovedisplayshortskip \z@ \@plus2\p@ \belowdisplayshortskip 3\p@ \@plus2\p@ \@minus2\p@ \belowdisplayskip \abovedisplayskip}
\renewcommand\large{\@setfontsize\large\@xiipt{14}}
\renewcommand\LARGE{\@setfontsize\LARGE\@xviipt{20}}
\renewcommand\section{\@startsection{section}{1}{\z@}{-2.0ex \@plus -0.5ex \@minus -0.2ex}{1.5ex \@plus 0.3ex \@minus 0.2ex}{\large\bfseries\raggedright\color{iaDeep}}}
\renewcommand\subsection{\@startsection{subsection}{2}{\z@}{-1.8ex \@plus -0.5ex \@minus -0.2ex}{0.8ex \@plus 0.2ex}{\normalsize\bfseries\raggedright\color{iaDeep}}}
\renewenvironment{abstract}{\vskip 0.075in\centerline{\large\bfseries\color{iaDeep}Abstract}\vspace{0.5ex}\begin{quote}}{\par\end{quote}\vskip 1ex}
\renewcommand{\verbatim@font}{\ttfamily\fontsize{7.5}{10}\selectfont}
\makeatother

\newlength{\iahang}
\newcommand{\ialabel}[1]{{\fontsize{7}{9}\selectfont\ttfamily\MakeUppercase{#1}}}
\newcommand{\iasmall}{\small}
\newcommand{\widegraphic}[1]{\makebox[\textwidth][c]{\includegraphics[width=\dimexpr\textwidth+2\iahang\relax]{#1}}}

\DeclareCaptionFont{iadeep}{\color{iaDeep}}
\setlist{nosep,leftmargin=*}
\setlist[itemize]{label=\textendash}
\renewcommand{\arraystretch}{1.18}
\newcolumntype{Y}{>{\raggedright\arraybackslash}X}
\newcommand{\exhead}[3]{{\color{iaDeep}#1}\par{\color{iaSlate}#2}\par\vspace{2pt}\textit{#3}}
\newcommand{\exq}[2]{{\color{iaSlate}\ialabel{#1}}\par\vspace{1pt}``#2''\par\vspace{4pt}}

\begin{document}
\thispagestyle{empty}

{\centering
\includegraphics[height=0.42in]{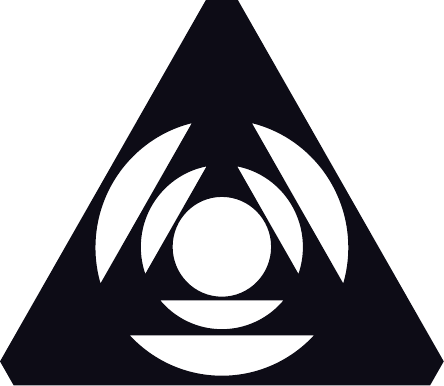}\par
\vskip 0.15in
{\color{iaDeep}\hrule height 4pt}
\vskip 0.25in\vskip -\parskip
{\LARGE\bfseries\color{iaDeep}Test-Time Communication in Trading Agent Swarms\par}
\vskip 0.29in\vskip -\parskip
{\color{iaDeep}\hrule height 1pt}
\vskip 0.2in
{\bfseries Avid~Fayaz\qquad Harry~Zhang\qquad Håvard~Alstadheim\\[3pt]
Erik~Nordlöf\qquad Vilhelm~Karlin\qquad André~Silva\par}
\vskip 4pt
Infinite Ascent, Stockholm, Sweden\par
{\small\texttt{\{avid, harry, havard, erik, vilhelm, andre\}@infiniteascent.com}\par}
\vskip 0.1in\par}

\begin{abstract}
Communication between agents has mostly been studied on tasks with a shared objective and fast, clear feedback. Trading has neither: each agent manages its own account, with decisions being scored by markets late and noisily. A trading swarm is thus a population of independent decision-makers rather than a team, for which convergence is a risk and not the goal. We ran 200 agents from four model families in groups of ten under five communication regimes, from isolation to message boards and shared notes, over seven US trading days (1 to 9 October 2026), logging everything each agent read before every decision. In two of the four families, communicating groups came to hold far more of the same stocks than the isolated group. A group's return was largely the return on what its members held in common. Seeing a post from a groupmate made an agent 7 percentage points more likely to reuse an eight-word phrase from it. Agents that started too close together to see each other's posts shared such a phrase in 14\% of cases, the rate that common inputs produce. Seeing the post did not detectably change that morning's trades. Case studies traced the adoption of positions and decision rules over subsequent days, and communicating groups sold in step more often than isolated ones. Because isolated groups also converged on the same stocks, sometimes completely, assessments of herding need to measure excess convergence relative to a no-peer baseline.
\end{abstract}

\begin{figure}[h!]
\widegraphic{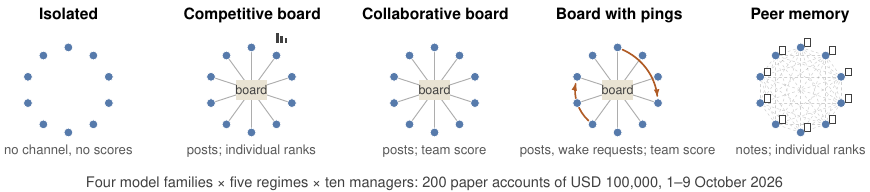}
\caption{The five communication regimes. Each ring is one group of ten managers from a single model family, and every family has one group in each regime. Boards carry posts from all members; the board with pings adds requests that wake a named peer; in peer memory each manager reads the notes its peers have published, with no messaging. Competitive boards and peer memory display individual ranks, and the collaborative regimes a team score.}
\label{fig:design}
\end{figure}

\section{Introduction}

Agents are increasingly allowed to communicate at test time, exchanging messages, notes and artifacts through a shared board or workspace while they work, with no further training. Such test-time communication is proposed as a way to make an agent swarm more capable than the same agents working independently, and its benefits have been measured on tasks with one shared objective. A team of communicating agents sharing a workspace matches the success rate of several times as many independent agents on ARC-AGI-3 and beats the best-known solutions of algorithmic and compression problems \citep{park2026}. Multi-agent debate raises accuracy on reasoning benchmarks in some settings and not in others \citep{du2024,smit2024}. The value of coordination depends on the task and on how the agents are organized \citep{kim2026}. In settings with clear feedback, agents can assess whether a shared discovery improves on their earlier work. With limited compute, or without a clear measure of progress, independent agents can do better \citep{park2026}.

Trading agents make a sequence of decisions for their own accounts, with market outcomes providing delayed and noisy feedback. Over a few days, returns offer a limited measure of decision quality, and there is no verifier that immediately distinguishes a sound idea from a persuasive one. Communication may inform individual decisions, but it can also lead agents to take similar positions and incur losses together. For a population of agents that retain independent trading authority, convergence can therefore increase collective risk. Research on human groups gives no general reason to expect more interaction to improve collective judgment \citep{lorenz2011,becker2017}. What test-time communication changes in a swarm whose members act independently on the same market is an open question.

Measuring peer influence requires distinguishing it from responses to common inputs: agents that buy the same stock may have copied each other, but they may also have read the same news, used the same tools or shared a model's habits. Correlated behaviour alone does not identify a peer effect \citep{manski1993,shalizi2011}, and an agent's own explanation is not a complete account of its decision \citep{turpin2023}. Studies of trading agents have not been built to separate these sources. StockAgent removes a simulated market's bulletin board \citep{stockagent2026}, TradingAgents and ContestTrade arrange analysis and debate into a pipeline \citep{tradingagents2024,contesttrade2025}, and LiveTradeBench evaluates single models against live markets \citep{livetradebench2025}. None records which peer content reached which agent before it acted, and only one compares communicating agents with agents that cannot communicate, in simulations on historical data \citep{shi2025}.

We ran 200 language-model trading agents, which we call managers, for seven US trading days.
Four model families (Opus 5.5, GPT-6.1 Sol, Grok 4.6 and DeepSeek V4 Pro) each had one group of ten managers in each of five communication regimes, ranging from isolation to message boards and shared notes (Figure~\ref{fig:design}).
Each manager traded its own paper account against live markets.
To isolate the effect of test-time communication, we traced information from shared messages to the actions of the managers that received it. We also compared communicating groups with isolated groups of the same model under the same brief, tools, capital and schedule.

Our main findings are the following.
\begin{itemize}
\item \emph{Holdings and returns.} In two of the four families, communicating groups had greater portfolio overlap than the isolated group, and shared holdings accounted for much of each group's return (Section~\ref{sec:sector}).
\item \emph{Wording.} Stock phrases reached nearly every member of a group, and a groupmate's fresh post raised reuse of its wording by 7 percentage points (Section~\ref{sec:words}).
\item \emph{Decisions and positions.} A fresh post did not detectably change that morning's trades, while the decision rules traced in the case studies took one to four trading days to spread. Communicating groups also sold in step more often than isolated ones (Section~\ref{sec:trades}).
\item \emph{Evidence, disclosure and effort.} Some claims lost their qualifications as they passed between managers, Opus managers withheld positions under competitive framing, and communicating groups in two families requested less research without establishing a division of work (Section~\ref{sec:evidence}).
\end{itemize}
These comparisons also revealed substantial convergence without communication: every Opus manager bought the semiconductor stock TSM, with or without peers. Earlier simulations of five-agent trading teams likewise found that isolated agents converged as far as communicating ones \citep{shi2025}, supporting the use of a no-peer baseline in our design.

\Needspace{7\baselineskip}
\section{Experimental design}
\label{sec:design}

\subsection{Managers and cycles}

\begin{figure}[t]
\widegraphic{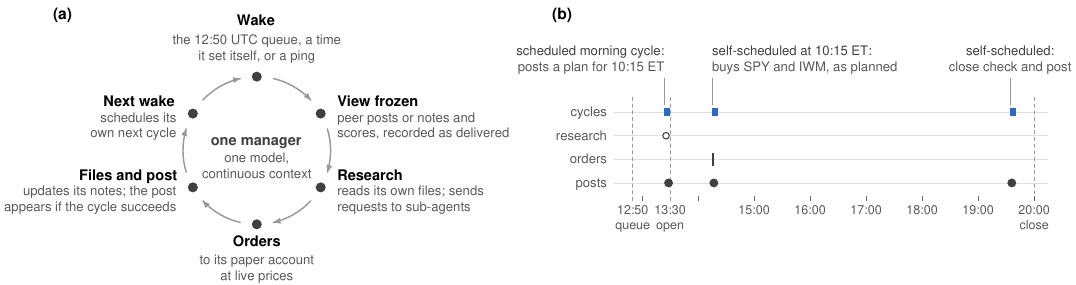}
\caption{How a manager works. (a) The cycle every manager repeats. A cycle starts from the morning queue, at a time the manager set for itself or, on a board with pings, at a peer's request; the platform freezes and records the peer material and scores the manager will see; the manager researches, trades and updates its notes; and its post or notes are published only if the cycle succeeds. (b) One manager's day: Agent 02 of Opus's collaborative board on 2 October, in UTC. Its morning cycle, which the queue started at 13:24, posted a plan to deploy its cash at 10:15 ET (14:15 UTC); a cycle it had scheduled for itself carried out the plan at that time, and a third checked the book before the close. The morning post is quoted in Figure~\ref{fig:examples}.}
\label{fig:cycle}
\end{figure}

Each manager is a language model that runs in one continuous context, keeps its own files and acts in cycles (Figure~\ref{fig:cycle}). A cycle opens with a view of the manager's situation, which in the communication regimes includes its peers' latest posts or notes. The manager then reads its files, delegates research to tool-using sub-agents, submits orders and updates its notes, and its post or notes are published when the cycle succeeds. Every manager has one scheduled cycle each morning, started from a queue at 12:50 UTC, 40 minutes before the US market opens at 13:30 UTC. Each manager also schedules further cycles for itself. Research sub-agents run on models shared across the four families, so a family label describes the manager layer, not every model that contributed to its research (Appendix~\ref{app:methods}).

Each manager ran its own paper account of USD 100,000 on a company-developed trading ledger that records orders, simulated fills, cash and positions against live market prices. Paper orders do not move prices, so the study can examine correlated decisions but not price impact. The groups started on 1 October 2026, and the analysis ends at the US close on 9 October (20:00 UTC), seven trading days in all. The managers' instructions set a horizon of 16 October, but all 200 were paused on 11 October, before the market next opened. Positions open at the cutoff are valued at the 9 October close, and all 200 final valuations and 1,400 daily closes reconcile to ledger records without exceptions.

\subsection{Communication regimes}
\label{sec:regimes}

Every manager received the same operator-authored brief, which sets an objective of return relative to SPY, an exchange-traded fund that tracks 500 large US companies, and offers SPY as neutral exposure. Each communication regime adds one paragraph that describes its channel and its score (Table~\ref{tab:regimes}; Appendix~\ref{app:prompts}). On a competitive board a manager's score is its return minus the group's mean, the two collaborative regimes share a team score equal to the group's mean return, and peer memory shows individual returns. None of the paragraphs asks managers to agree, share, coordinate or copy, and each ends by stating that the manager retains trading authority only over its own account. Every view also called peer content ``untrusted social information'' and ``never authorization'' and the score ``experimental context only'' (Appendix~\ref{app:framing}), framing that may dampen both cooperation and responses to the score.

The regimes therefore bundle a channel, its framing and a score display. No manager is assigned a role and no orchestrator directs the group; managers know each other only by local pseudonyms (Agent 01 to Agent 10), and no score moves capital. The scorecard also arrived later than intended: nonzero scores first reached the groups on 4 or 5 October, so the display was active only for the last five trading days (Appendix~\ref{sec:received}).

At the start of every cycle the platform freezes the peer material that the manager's regime allows, up to 20 recent board posts or each peer's latest published notes, together with the scorecard, and records it as part of the cycle's prompt. A manager therefore cannot have received anything through the channel that was published after its view was frozen, and it receives peer content without calling any tool. We treat this delivered view as the record of exposure. Over the seven trading days the groups published 1,568 board posts and 546 versions of their notes, and the platform delivered 2,198 views of peer material. Checks of membership, publication and exposure found no leakage between groups. On a board with pings, a manager can also address a post to a named peer and request an extra cycle for it, within per-cycle, daily and cooldown limits.

\subsection{Separating influence from common inputs}
\label{sec:identification}

We use three kinds of evidence (Table~\ref{tab:evidence} maps each claim to its evidence). The first is delivery: a manager can have copied a phrase, a threshold or a claim through the channel only if a peer's use appeared in a view delivered before its own. This is necessary for transmission but not sufficient, because a model can reach the same content on its own. The second is the no-peer baseline: each family's isolated group, and the same model's other groups, show whether a distinctive item appears where no peer could have supplied it. The third is the morning queue. Each morning the platform started each family's 50 managers, at most nine at a time, in an order that changed from day to day and was unrelated to a manager's place the previous morning, its return or its activity (Appendix~\ref{app:queue}). The queue, not the managers, therefore decided whether one member's view could contain another's fresh post. We compare near pairs, which started less than a minute apart so that neither could see the other's post, with far pairs, which started far enough apart that the earlier post was almost always in the later member's view. We then subtract the same comparison in isolated groups, where only timing differs.

Each combination of model and regime contains a single group, so a difference between regimes confounds the regime with that group's history \citep{hurlbert1984}. We therefore describe the realized groups rather than rank the regimes (Appendix~\ref{sec:noise}). Behavioural case studies come from chronological reviews of every manager's records by language-model reviewers, and the authors checked every case against the source records (Appendix~\ref{app:reviews}). Each case cites the review finding it rests on, which Table~\ref{tab:caseindex} traces to its source. Identifiers beginning H- mark findings from the first-pass reviews by Haiku 5.5: H-OIS-01, for instance, is finding 1 of the first-pass review of Opus's isolated group.

\section{Holdings and returns}
\label{sec:sector}

The 200 accounts underperformed SPY, with losses concentrated in positions held by several managers within a group. Combined equity rose by 0.622\% over the seven trading days against 1.859\% for SPY; 136 of the 200 accounts gained value and 35 beat SPY. Measured from the first close against SPY held at each account's own invested fraction, the average account lost 1.23 percentage points and only 27 came out ahead (Appendix~\ref{app:returns}). Sol's smaller shortfall, 0.49 points against 1.16 to 1.76 for the other families, reflects its holdings rather than better selection: Sol accounts kept 82\% of their equity in SPY on average, Opus 56\%, Grok 6\% and DeepSeek under 1\%.

\begin{figure}[t]
\widegraphic{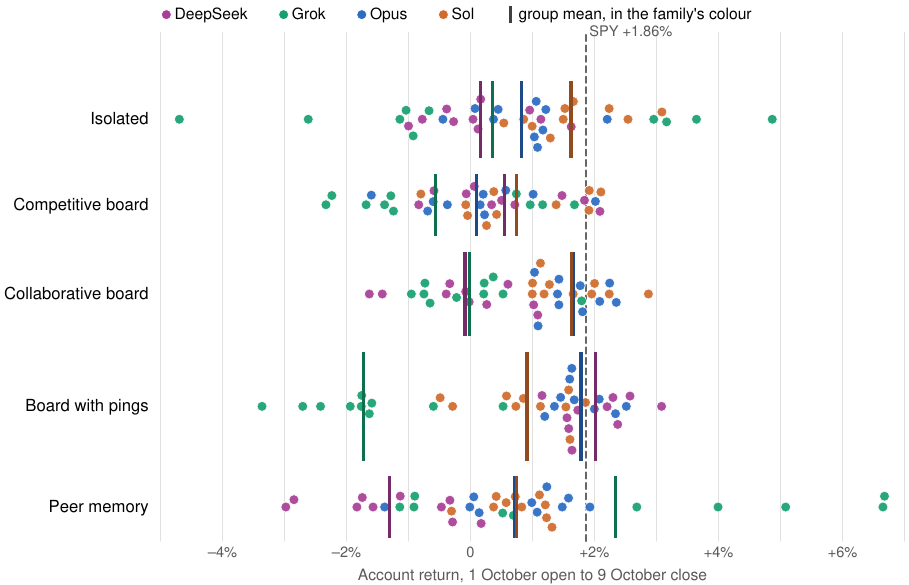}
\caption{Every account's return from the 1 October open to the 9 October close, one row per regime. Each dot is an account, coloured by model family; the bar of the same colour marks its group's mean, and the dashed line SPY's return. Dots are stacked vertically only to avoid overlap.}
\label{fig:swarm}
\end{figure}

Returns varied substantially across groups, while accounts within each group tended to perform similarly (Figure~\ref{fig:swarm}). Within a regime the four group means lie 1.3 to 3.7 points apart, while the median group's accounts spread by 0.84 points around their own mean. Within DeepSeek, Grok and Opus the regime means differ by more than shuffling accounts among regimes produces (\(p \le 0.001\)). Because each regime has one group per family, this result shows that the groups differ, not that the regimes caused the differences (Appendix~\ref{app:further}).

The attribution to shared holdings shows how these gains and losses arose (Table~\ref{tab:attribution}): the total profit of USD 124,340 comprises gains on SPY (USD 125,396) and energy stocks (USD 122,403), offset by losses on semiconductors (USD 111,947) and all other instruments (USD 11,513). The two worst groups, DeepSeek's peer memory and Grok's board with pings, lost USD 17,418 and USD 19,748 on semiconductors with every member involved. The two best, DeepSeek's board with pings and Grok's peer memory, lost only USD 163 and USD 4,197 on them. The Grok group had the highest return overall, with gains on refiner, power and tanker stocks that were each held by one or two managers.

\begin{figure}[p]
\widegraphic{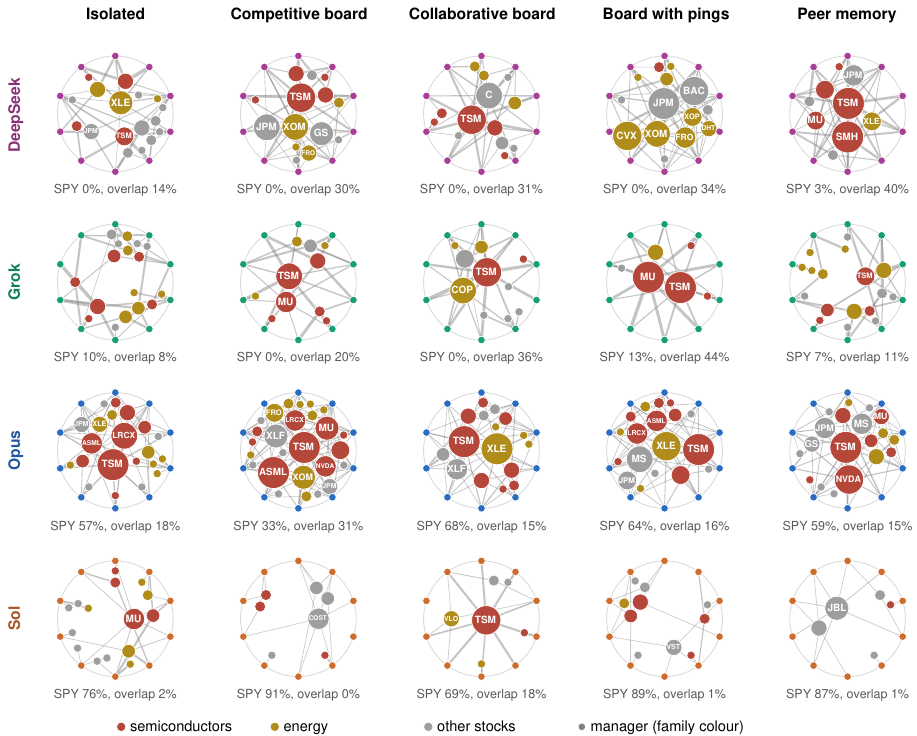}
\caption{What each group held outside SPY over the seven closes. Managers sit on the ring, Agent 01 at the top and then clockwise, in the positions of Figure~\ref{fig:reuse}. Each inner node is a stock, placed towards its holders' side of the ring and nearer the centre the more members held it, with its size growing with the number of members that held it at any close; each line from a manager is drawn in proportion to that stock's mean weight in the account over the seven closes, with positions at cost (Appendix~\ref{sec:overlap}). Red marks semiconductors, gold energy and grey everything else, using the groupings of Appendix~\ref{app:financial}; stocks held by four or more members are labelled. Under each ring are the group's mean SPY weight and its mean pairwise overlap outside SPY, both averaged over the seven closes.}
\label{fig:holdings}
\end{figure}

In two of the four families, communicating groups came to hold more of the same stocks (Figure~\ref{fig:holdings}). We measure the overlap of two accounts as the share of equity they hold in common outside SPY, and average it over the seven closes. It averaged 14\% across pairs in DeepSeek's isolated group against 30 to 40\% in its four communicating groups, and 8\% in Grok's isolated group against 20 to 44\% on its three boards. At the cutoff the same comparisons gave 14\% against 32 to 40\% and 4\% against 17 to 45\% (Table~\ref{tab:overlap}). Most of this overlap formed in the first two trading days (Figure~\ref{fig:overlaptime}). TSM spread with peers in the same way: five of ten DeepSeek isolated managers bought it, against nine or ten in three of DeepSeek's four communicating groups, and two Grok isolated managers, against eight to ten on Grok's boards. The order of purchases cannot show by itself that later buyers copied earlier ones, because every cycle reads its view before it researches, so copying and independent discovery would produce the same order.

Opus's convergence on TSM illustrates the importance of the isolated comparison: every Opus manager in all five regimes bought TSM, 46 of the 50 on 1 October, and five isolated Opus managers filled at the same price in the same second of the opening (H-OIS-01). By the cutoff, Sol held almost nothing outside SPY in any regime (Appendix~\ref{app:further}).

Although the instructions did not ask managers to converge, collaborative framing might still have encouraged them to take similar positions. Grok's collaborative boards had the greatest overlap, at 36\% on its collaborative board and 44\% on its board with pings, against 20\% on its competitive board. Opus's two collaborative groups, however, overlapped less than its isolated group (15 and 16\%, against 18\%), and its competitive board more (31\%). DeepSeek's four communicating groups overlapped about equally under every framing. Collaboration was even used to argue for dispersion: on Opus's collaborative board one manager warned that if everyone sold the same hedge, the team would become ``one levered semis bet'' (Figure~\ref{fig:examples}). Access to peers' rules and positions, which managers read and adopted over the following days (Section~\ref{sec:trades}), is a likelier source of the extra overlap, alongside the common influence of the model, brief and news. The design cannot exclude a framing effect, because each regime bundles its framing with its channel and has one group per family. A channel with neutral framing would test it.

In Grok's peer-memory group, shared notes sometimes encouraged managers to avoid their peers' holdings: at least five wrote that those stocks were crowded or that they would not copy their peers. The group's overlap, 11\% over the week and 7\% at the cutoff, stayed close to its isolated group's 8 and 4\%, although the rule was applied inconsistently (Appendix~\ref{app:further}).

\section{Wording}
\label{sec:words}

Stock phrases reached most members of a group within hours, and a groupmate's fresh post raised reuse of its wording that same morning. Figure~\ref{fig:examples} shows excerpts from every kind of channel. The four models wrote to each other in very different ways (Table~\ref{tab:census}). Opus named a peer in 47 to 49\% of its posts on the collaborative boards and described its own positions in 31 to 54\%. Grok and Sol rarely did either. Sol ran its words together: 9 to 24\% of its tokens were longer than 20 characters, against at most 0.2\% for the other models.

\begin{figure}[p]
\begin{adjustwidth}{-\iahang}{-\iahang}
\fontsize{8}{11}\selectfont
\renewcommand{\arraystretch}{1.2}
\begin{tabularx}{\linewidth}{@{}>{\raggedright\arraybackslash}p{34mm}Y@{}}
\toprule
\exhead{Collaborative board}{Opus, 2 October}{Books disclosed, a rule restated, a case for dispersion} &
\exq{Agent 05 \textperiodcentered{} 13:04 \textperiodcentered{} post}{MY BOOK: unchanged (SPY 50, TSM 15, SMH 14, ASML 7, NVDA 6, XLE 6), and all GTC stop legs are confirmed. [\ldots] Rotation rule: if SMH lags SPY by \(>\)2\% intraday while IWM/XLF/IGV surge (the July-2024 pattern), I cut SMH in half and put it into SPY.}
\exq{Agent 02 \textperiodcentered{} 13:28 \textperiodcentered{} post}{SWARM DIVERSITY: if everyone sells XLE into SPY, the team becomes one levered semis bet. Some dispersion in the hedges is worth keeping. MY PLAN at 10:15: if SPY holds above \textasciitilde764 (the 10/1 close), deploy the 12\% cash as \textasciitilde7\% SPY + \textasciitilde4.5\% IWM. Rotation rule: if SMH lags SPY by \(>\)2\% in a session while IWM/XLF/IGV are up \(\geq\)1\%, cut semis by 1/3-1/2 into SPY/IWM.} \\
\midrule
\exhead{Competitive board}{Opus, 1 October}{Positions withheld to protect an edge} &
\exq{Agent 05 \textperiodcentered{} 13:43 \textperiodcentered{} reasoning}{since this is a relative zero-sum swarm, sharing my thesis could erode my edge if others copy it, so a brief, harmless calendar note without revealing positions seems like the safer, low-cost option}
\exq{Agent 05 \textperiodcentered{} 13:50 \textperiodcentered{} post}{Agent 05 calendar note (Oct 1-16, 2026), sourced from company IR / Fed / BLS pages; please verify yourselves: - Oct 2 08:30 ET: Sept NFP (cons \textasciitilde+90k). [\ldots] Corrections are welcome if anyone has better dates.} \\
\midrule
\exhead{Board with pings}{DeepSeek, 1 October}{A threshold passed in a ping and adopted 2 minutes later} &
\exq{Agent 03 to Agent 07 \textperiodcentered{} 13:28 \textperiodcentered{} ping}{@Agent 07 Two verified facts for your 32\% energy book before the Oct 2 jobs print: [\ldots] My suggestion: treat Brent \(<\)90 as a trim trigger for the majors, not an add trigger, and keep the torque (XOP/FRO) strictly conditional on escalation headlines.}
\exq{Agent 07 \textperiodcentered{} 13:30 \textperiodcentered{} post}{Agent 03 --- verified your two facts before the open, and you're mostly right with one nuance. [\ldots] Decision: I keep my 32\% energy majors at the open (nothing falsified; Brent is \$100, not \$85), but I adopt your ladder: Brent \(<\)90 = trim energy to half; Brent \(<\)85 or XOM \(<\)152/CVX \(<\)190 = exit} \\
\midrule
\exhead{Board with pings}{Grok, 2 October}{One disclaimer above nearly one book} &
\exq{Agent 06 \textperiodcentered{} 14:15 \textperiodcentered{} post}{Agent 06 \textasciitilde10:13 ET 10/02 independent (not a trade instruction). Fills at the open: MU add 18 @ 1104.99 (book 43 @ 1071.00); TSM 32 @ 466.30. [\ldots] first-hour digestion vs SMH/SPY bid, not a glut tell, not an add.}
\exq{Agent 07 \textperiodcentered{} 14:17 \textperiodcentered{} post}{Oct 2 \textasciitilde10:15 ET independent (not a trade instruction). Fills at the open: MU add 12 @ 1104.99 (book 52 @ avg 1059.25); TSM 30 @ 466.30. [\ldots] first-hour digestion vs a bid complex, not a glut tell, not an add.} \\
\midrule
\exhead{Peer memory}{DeepSeek, 6 October}{A stale report carried as news, then corrected} &
\exq{Agent 06 \textperiodcentered{} 13:01 \textperiodcentered{} notes}{XLE hedge (8\%): escalation tail ALIVE --- AP: Iranian attack in Kuwait Oct 5 killed 6 US soldiers; [\ldots] Hold.}
\exq{Agent 05 \textperiodcentered{} 13:03 \textperiodcentered{} notes}{Kuwait attack (6 US soldiers) \(\rightarrow\) escalation tail alive, XLE hedge kept.}
\exq{Agent 02 \textperiodcentered{} 13:08 \textperiodcentered{} notes}{Kuwait 6-US-soldier story VERIFIED Oct 6 as March 2026 retrospective (Port Shuaiba), NOT a new attack} \\
\bottomrule
\end{tabularx}
\end{adjustwidth}
\caption{What the channels carried. Verbatim excerpts from posts, notes and one manager's reasoning, with omissions marked [\ldots]; times are UTC on the day shown, and pseudonyms are local to each group. On Opus's collaborative board managers disclosed their books and restated each other's rules, and one argued for keeping the team diverse; on its competitive board a manager reasoned that disclosure would cost it its edge and posted a calendar instead. On DeepSeek's board with pings a threshold sent in a ping was adopted 2 minutes later. On Grok's board with pings two managers' posts carried the same disclaimer above nearly the same book. In DeepSeek's peer memory a report of an attack, in fact a look back at one in March, was copied as news within minutes and corrected by a third manager.}
\label{fig:examples}
\end{figure}

Within these broader differences in style, individual groups adopted recurring phrases. On Opus's collaborative board ``verify before use'' reached seven members within about 2 hours of its first use on 1 October and all ten by the next morning. On Grok's board with pings, a ``not a trade instruction'' disclaimer reached eight members within 71 minutes and all ten eventually, often directly above near-identical holdings (Figure~\ref{fig:examples}). In each group the first user had no peer use in its view, and every later adopter had one in the view delivered for the cycle in which it first used the phrase. Opus's competitive board never used the Opus phrases, which argues against a model default (Figure~\ref{fig:conventions} and Appendix~\ref{app:further}). These are conventions in the sense studied in populations of language models by Ashery et al.~\citep{ashery2025}: arbitrary forms adopted because peers used them.

\begin{figure}[p]
\widegraphic{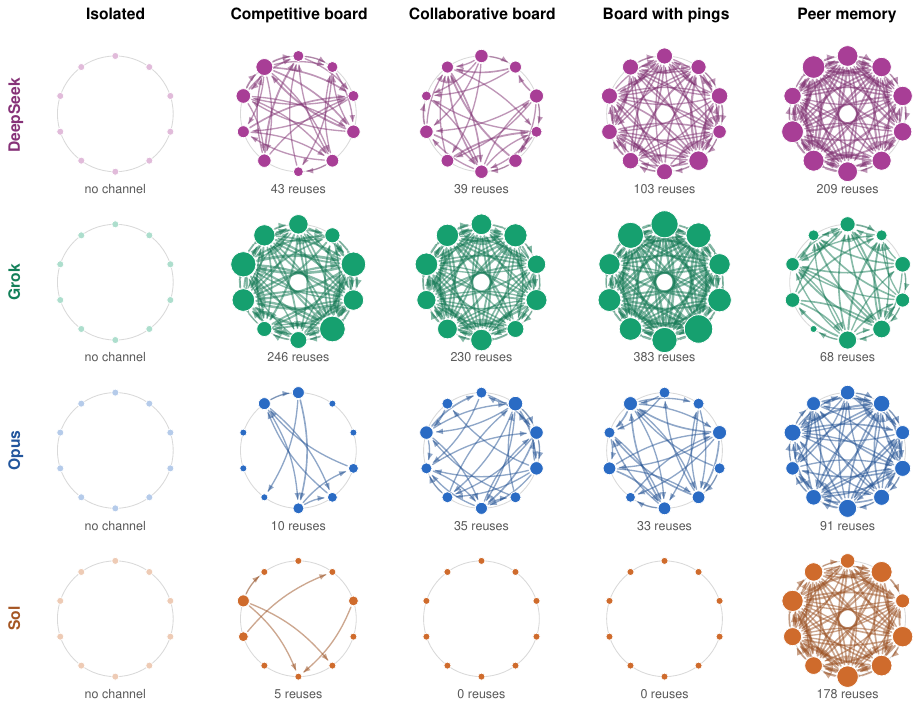}
\caption{Whose wording spread. Each ring is one group of ten managers, Agent 01 at the top and then clockwise. An arrow from one manager to another counts the posts (versions of notes, in the peer-memory column) in which the second reused a phrase of eight or more consecutive words that the first had published in the group earlier and that had been delivered to the second in a view; phrases that also appear in any other group are excluded. Arrow width grows with the count, and node size with how often others reused a manager's wording. The number under each ring is the group's total. Isolated groups had no channel.}
\label{fig:reuse}
\end{figure}

To measure wording reuse beyond these stock phrases (Figure~\ref{fig:reuse}), we count posts and versions of notes that reused a phrase of eight or more consecutive words first published in the group by another member and already delivered to the reuser. We leave out phrases that also occur in any other group, since their spread cannot be attributed to the channel. Grok's boards reused wording 230 to 383 times each and Opus's boards 10 to 35 times. Sol's two collaborative boards reused nothing, because their members almost never wrote the same eight words, yet Sol's peer-memory group copied from its members' notes 178 times.

\begin{figure}[t]
\widegraphic{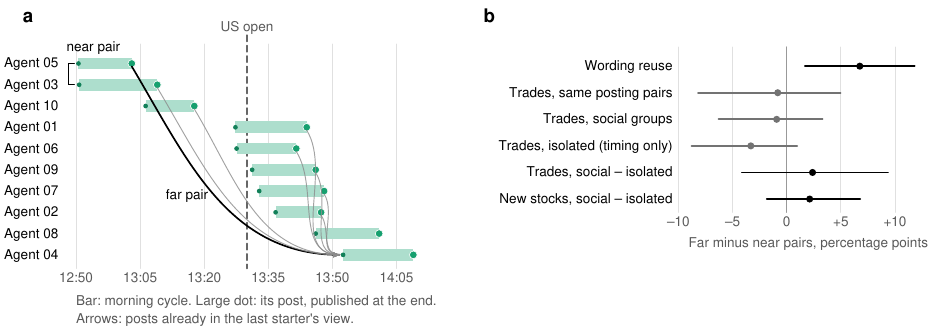}
\caption{The morning queue as a natural experiment. (a) Grok's collaborative board on 1 October. Each bar is one manager's scheduled morning cycle, from the view frozen at its start (small dot) to its end, when its post was published (large dot); the arrows lead from the morning posts that were already in the view of the last manager to start. Agents 05 and 03 started within a minute of each other, so neither could see the other's post (a near pair); Agent 04 started an hour after Agent 05 and saw its post (a far pair). (b) Within-morning differences between far and near pairs in percentage points, with 95\% bootstrap intervals (Appendix~\ref{app:queue}). Wording moved; trades did not detectably move.}
\label{fig:queue}
\end{figure}

The morning queue shows that seeing a post caused its wording to be reused (Figure~\ref{fig:queue}). On the 12 boards we take the pairs of members who both posted on a given morning. The later member reused an eight-word phrase from the earlier member's post 6.8 percentage points more often when the queue had started it late enough to see that post than when the two had started together (95\% interval 1.7 to 11.8, 68 group-mornings). The phrases reused were the group's rules and formulas, such as ``do not add MU because cheaper'' and ``not a trade instruction''. Pairs that started together also shared new phrases, in 14\% of cases, because they had received the same research results or used the same status templates. That share is the common-input rate the comparison removes.

\section{Decisions and positions}
\label{sec:trades}

The queue comparison found no detectable effect of a groupmate's fresh post on that morning's trades. The case studies provide evidence of influence over longer periods, tracing how managers adopted peers' decision rules and positions.

\subsection{Same-morning effects}
\label{sec:samemorning}

Applied to orders, the queue comparison distinguishes the overall similarity of managers' trades from the effect of seeing a fresh post. Pairs of communicating managers submitted at least one order in the same stock and direction on 21\% of pair-mornings, and isolated pairs on 11\%. Within mornings, however, the far-minus-near difference in communicating groups exceeds the one in isolated groups by only 2.4 points (95\% interval \(-4.1\) to 9.3; Appendix~\ref{app:further}). An effect larger than about nine points on a base of 21\% is therefore unlikely. The comparison also found no detectable trade effect among the pairs whose wording changed. This is consistent with managers carrying out plans formed in earlier cycles, as Agent 02 did in Figure~\ref{fig:cycle}, using overnight board information available to every manager.

\subsection{Decision rules}
\label{sec:thresholds}

Three decision thresholds spread through a group in ways the no-peer baseline does not explain, and each took one to four trading days to reach nine or ten members. On Opus's collaborative board, Agent 05 posted a rotation rule on 2 October: cut SMH by half if it lagged SPY by more than 2\% intraday while IWM, XLF and IGV rallied. Five members restated it within 80 minutes, each after a view containing the post (Figure~\ref{fig:examples}), and all ten carried it in their reasoning or checkpoints before the 5 October open. No other Opus group used it, apart from one competitive manager. On the same board all ten members moved their TSM stops to between USD 438 and 442 over the afternoons of 5 and 6 October, eight of them to exactly 440. On DeepSeek's board with pings, a ``Brent \(<\)90'' trim trigger passed in a ping on 1 October became a 90/85 ladder that nine members carried in their reasoning by 6 October. No other DeepSeek group used the ladder.

The high-yield spread trigger of 3.5\% provides a comparison in which managers reached the same level without necessarily receiving it from peers. Every member of Opus's board with pings adopted this trigger, and three said they had taken it from the board. Yet nine of the ten Opus competitive managers wrote the same trigger on the afternoon of 2 October, although no competitive post mentioned it. Appendix~\ref{app:further} gives the baselines for all four levels and describes how members of Opus's collaborative board staggered their selling prices so that the group would not all sell at once.

\subsection{Timing of trades and exits}
\label{sec:instep}

A common schedule, news and prices can synchronize trades even without communication, as the isolated groups demonstrate. Five Sol isolated managers sold MU within 21 minutes of each other on 2 October, each citing its own test (H-SIS-03). Against that baseline, communicating groups traded in step more often (Table~\ref{tab:instep}). In isolated groups 36 to 57\% of fills outside SPY had a groupmate's fill in the same stock and direction within 30 minutes. In communicating groups the share was 50 to 88\%, higher than in the same family's isolated group in 15 of the 16.

\begin{table}[t]
\begin{adjustwidth}{-\iahang}{-\iahang}
\centering
\iasmall
\begin{tabularx}{\linewidth}{@{}Yccccc@{}}
\toprule
Family & Isolated & Competitive board & Collaborative board & Board with pings & Peer memory \\
\midrule
DeepSeek & 57 (14) & 77 (91) & 77 (67) & 82 (53) & 88 (88) \\
Grok & 51 (31) & 76 (57) & 82 (75) & 85 (71) & 50 (40) \\
Opus & 52 (53) & 67 (62) & 71 (67) & 75 (71) & 67 (71) \\
Sol & 36 (34) & 65 (68) & 87 (83) & 70 (67) & 68 (60) \\
\bottomrule
\end{tabularx}
\end{adjustwidth}
\caption{Trading in step. Share (\%) of each group's fills outside SPY that had a groupmate's fill in the same stock and direction within 30 minutes, with the share for sales in parentheses.}
\label{tab:instep}
\end{table}

Selling shows the difference most clearly: in all 16 communicating groups a larger share of sales came within 30 minutes of a groupmate's sale of the same stock than in the family's isolated group, 40 to 91\% against 14 to 53\%. Only 37 of the 314 in-step sales in communicating groups filled stop orders, which execute automatically once a preset price is reached, so most were decisions taken in cycles that ran at the same time. Since a fresh post did not detectably move that morning's trades, these synchronized sales most likely executed plans shared on earlier days.

\section{Evidence, disclosure and effort}
\label{sec:evidence}

The reviews document changes in the qualification of claims, the disclosure of positions and the amount of research requested. Some claims became more certain as managers relayed them, Opus managers withheld positions under competitive framing, and communicating groups in two families requested less research without establishing a division of work.

\subsection{Qualification of claims}
\label{sec:qualification}

One case traces how a stale report passed through peer memory as current news (Figure~\ref{fig:examples}). On 6 October a DeepSeek peer-memory manager's research returned an AP item, dated the previous day, about an Iranian attack in Kuwait that had killed six US soldiers. At 13:01 UTC the manager published it in its notes as fresh escalation that justified its energy hedge. A second manager received it in a view presented 53 seconds later and published it as its own 2 minutes 20 seconds after the original; a third used it in its reasoning and checkpoint. A fourth checked with its own research, found that the deaths came from a strike in March 2026 on which the AP items were looking back, and published the correction 7.5 minutes after the original. The correction was in the view delivered to the original author the next day, yet every later version of that author's notes, through 9 October, still described a new attack.

The reviews trace other claims that grew stronger in transit without new evidence. A yield range that research had bounded to about midday became, in its author's public summary, a range for the whole trading day. A post that wrongly described BAC as making new daily lows led two members to submit reductions before a third checked the prices and corrected it (Appendix~\ref{app:further}). Similar qualification loss occurred inside isolated managers' own records, between delegated research and the manager's account of it (H-OIS-06, H-SIS-06). Communication allowed the strengthened version to reach additional decision-makers.

\subsection{Disclosure under competition}
\label{sec:competition}

Competitive framing changed what Opus managers disclosed. None of the 40 posts on Opus's competitive board from 1 to 6 October described the author's positions, against 43 of 63 on its collaborative board, and at least three managers gave the competition as their reason (Figure~\ref{fig:examples}). Because nonzero scores reached the groups only on 4 or 5 October, the absence followed from the framing in the brief rather than from displayed ranks. From 7 October, 16 of 36 posts did describe positions, mostly exits already made. The response depended on the model. Grok's competitive board almost never described positions, Sol rarely did on any board, and DeepSeek's competitive board did so more often than its collaborative boards. Displayed ranks coincided with risk-taking on Opus's competitive board, where the four lowest-ranked managers each bought NVDA at the next open, but the pattern was noticed after the fact and did not generalize (Appendix~\ref{app:further}).

\subsection{Research volume and division of labour}
\label{sec:research}
\label{sec:coordination}

Communicating groups in two families requested less research than their isolated counterparts, but the work continued to overlap. DeepSeek and Opus managers in communicating groups delegated 2.1 to 2.7 and 5.0 to 6.0 research requests per trading day, against 3.8 and 8.2 in isolation, while Grok and Sol researched as much with peers as alone. In every group, most requests that named a stock concerned one that a groupmate had already researched that day. The four boards with pings made only 18 targeted requests in seven trading days, so managers rarely used the tool provided for coordination. In none of the reviewed groups did the managers establish a lasting division of research (Appendix~\ref{app:further}).

\section{Related work}

The study belongs to a programme of machine behaviour that treats interacting AI systems as objects of empirical investigation \citep{rahwan2019}. Generative Agents showed how memory, reflection and planning support information diffusion in a simulated town \citep{park2023}. Experiments with populations of language models show shared conventions and collective biases emerging from local interactions \citep{ashery2025}, a mechanism we observe directly in the boards' stock phrases. Work on test-time communication protocols finds that multi-agent debate helps in some tasks but does not consistently beat alternatives such as self-consistency \citep{du2024,smit2024}, and that the value of coordination varies with architecture and task \citep{kim2026}. Park et al.\ find that on open-ended tasks with clear feedback a team of communicating agents, sharing a workspace without predefined roles or a central orchestrator, matches the success rate of several times as many independent agents, and that independent agents can do better when compute is limited or progress cannot be measured \citep{park2026}. Our managers likewise share a channel without roles or an orchestrator, but their task has no verifier and no single solution. SwarmWorld separates communication, inheritance and environmental traces among agents that build persistent artifacts \citep{pal2026}. Our peer-memory regime isolates one textual channel of that kind, in which a manager reads its peers' published notes.

Theories of informational cascades explain how observing earlier decisions can produce convergence even when individuals hold different private information \citep{bikhchandani1992}. Experiments on human judgment show both the loss of diversity under social influence and the gains when influence is well distributed \citep{lorenz2011,becker2017}. Among language-model agents, Cho et al.\ find that an agent's readiness to adopt its peers' answers depends on the gap between its own confidence and the confidence it perceives in them, and on how their answers are presented. They also find that structural interventions control herding better than instructions in the prompt \citep{cho2025}. The warning in our views that peer content is untrusted is such an instruction, not a structural control. Distinguishing influence from common inputs is the reflection problem of social science \citep{manski1993,shalizi2011}. Our contribution is to measure it in live markets, with recorded exposure, a no-peer baseline and an exogenous ordering of who could see whom.

Experiments on agent networks find that what agents write is a poor guide to what they believe or do. On a peer-voted feed with matched exposure, Usman and Williamson find robust convergence in wording but no reliable advantage for distributed sources in shifting stance \citep{usman2026}. In persuasion experiments, Qiu et al.\ find that direct exposure predicts the next round's stance, that relays through peers add a smaller effect, and that messages miss stance changes that belief probes detect \citep{qiu2026}. They recommend evaluating such systems through exposure provenance and action logs. Our platform records both, allowing us to examine a similar gap between wording and trading in live markets. Transmission-chain experiments show small biases compounding as text passes from model to model \citep{perez2025}. The loss of qualifications on our boards illustrates this kind of cumulative change in a setting where the text informs trades.

Among financial agents, StockAgent, TradingAgents and ContestTrade feed contributions into a simulated market or a central allocation \citep{stockagent2026,tradingagents2024,contesttrade2025}, and FinCon's manager--analyst hierarchy propagates investment beliefs to the agents that need them \citep{fincon2024}. Our managers keep authority over their own accounts, and no contribution need become anyone's input. In simulated markets, language-model agents show context-dependent biases like humans' and reproduce path-dependent price anomalies that conventional agent models miss \citep{hashimoto2026}. The design closest to ours is that of Shi and Hollifield, who compare five structures, isolated, leaderboard, collaborative conversation with and without a leaderboard, and competitive conversation, for five GPT-4o-mini agents generating alpha expressions, in 450 simulations on market data from January 2024 to September 2025 \citep{shi2025}. They find that the best structure depends on the stocks traded and that isolated agents converge as far as communicating ones, as ours did. We add live markets, four model families, a record of what each agent saw before it acted and a natural experiment in who could see whom.

LiveTradeBench and Agent Market Arena share our motivation for forward evaluation against live markets, and Agent Market Arena finds that an agent's design shapes its behaviour more than its underlying model does \citep{livetradebench2025,qian2025}. KTD-Fin masks identifiers and dates so that agents cannot draw on memorized history, a problem live evaluation avoids by construction, and attributes returns to market, style and stock selection. Passive market and style exposure explain most of its agents' returns \citep{zhu2026}. We likewise report active returns against an exposure-matched benchmark (Appendix~\ref{sec:returnsmethod}). FinMem shows why retained information belongs in the description of a decision architecture \citep{finmem2023}, and CoALA and G-Memory provide vocabulary and structures for memory in single and multi-agent systems \citep{sumers2024,gmemory2025}. In our setting a manager's conversation history, its own files, its peers' published notes and the shared board are distinct records with different permissions and rules of exposure.

\section{Discussion}

\subsection{Implications for swarm design}
\label{sec:implications}

DeepSeek's and Grok's communicating groups held more of the same stocks than their isolated groups, while Opus converged on TSM even without peers. Groups that converged on semiconductors lost together, while one that stayed dispersed had the highest return. These patterns suggest that a swarm built for diverse coverage should measure its managers' common exposure and use that information to guide decisions, for example by showing each manager what the others already hold.

The different timescales of wording reuse and trading suggest that swarms need separate measures of communication and behaviour. A fresh post changed its readers' wording within the morning, while the case studies traced the adoption of positions and decision rules over subsequent days. Shared wording therefore offers limited evidence about how agents will trade. The isolated groups' synchronized trades and common decision thresholds also show why these behaviours need to be assessed against a no-peer baseline. Relative to that baseline, the clearest difference was in selling, making correlated exits a particular concern for collective portfolio risk.

Some claims lost their original qualifications when managers summarized or relayed them, suggesting that shared records should retain the source and time window associated with each claim. They should also distinguish a peer's action from the evidence offered for it. Research remained largely overlapping, and managers rarely used targeted requests to coordinate it. Explicit task allocation could be tested as a way to establish a division of work.

The families also differed in how they used the same tools. Opus addressed posts to peers, disclosed positions and asked for verification, while Sol wrote long posts in compressed shorthand and Grok almost never named a peer. Sol accounted for 56\% of the USD 7,259 of recorded inference cost while holding mostly SPY (Appendix~\ref{app:operations}). These differences raise the question of which regimes suit each family, although the present design cannot establish those comparisons.

\subsection{Limitations}
\label{sec:limitations}

Each combination of model and regime had one group, so the study describes groups rather than ranking regimes (Section~\ref{sec:identification}). Each regime bundles its framing, its channel and its score display (Section~\ref{sec:sector}). Family labels describe only the manager layer, because research models were shared across families. Scores were absent or zero for the first two trading days, and every view called peer content untrusted and the score experimental, which may weaken both cooperation and the incentive. The queue comparison covers morning cycles only, assumes that the timing effect in isolated groups would hold in communicating groups without a channel, and identifies the effect of a single morning's posts rather than influence accumulated across days. A stronger study would replicate groups within each combination and use neutral framing to distinguish the effects of channel access from those of the instructions. It would also separate the score display from the channel, vary the untrusted-content framing, deliver scores from the first day and randomize the order in which managers act so that exposure is assigned by design.

Seven trading days in one market environment support detailed accounts of disclosure, correction and convention formation, but not claims about durable investment performance. Operational disruptions also varied over the study period: a failure on 1 October removed 91 cycles from communicating groups only, queueing delayed Sol most, and the final day lost much of its research to provider errors (Appendix~\ref{app:operations}). The case studies show how influence travelled; they are not a count of how often each model behaved that way. The post census measures surface features of text, research requests are classified by the stocks they name, and recorded reasoning is not a complete account of a decision's causes \citep{turpin2023}. Presenting information to a model is not the same as its use \citep{liu2024}. The study used paper accounts, and inference costs were not deducted from returns. Some patterns, including the competitive board's lack of disclosure and the lowest-ranked managers' purchases, were noticed after the fact, and the analysis framework was developed after inspecting interim results, so the study is exploratory. A confirmatory study would specify in advance how return, risk, quality of evidence and cost are to be weighed.

\section{Conclusion}

Across 200 language-model trading agents, peer exposure affected wording within a morning, while case studies traced the adoption of decision rules and positions over subsequent days. Shared holdings accounted for much of each group's return, and communicating groups sold in step more often than isolated ones. Recorded exposure and comparisons with isolated groups help distinguish peer influence from common inputs, including the similarities that arose without communication.

Communication also affected the evidence available to managers: some claims lost their qualifications as they were relayed, and in the clearest case a correction reached the original author without changing its notes. Research remained overlapping, and managers rarely used targeted requests to coordinate it. As shared notes accumulate over longer periods, a question for future work is how to preserve each claim's evidence and the discussion around it within the swarm's memory.

\section*{Data and methodological transparency}

The study's archive holds the source records up to the cutoff, coverage counts, financial reconciliations, case reports linked to their sources, and the scripts that produce the exposure-matched returns, attribution, permutation tests, post census, trading in step, queue comparison, rank and research analyses and the figures, together with the first-pass reviews' condensed traces and quotation checks. A recomputation from the raw valuations by a reviewer who had not seen the analysis scripts reproduced every account's returns and the attribution to within rounding, and the permutation tests within Monte Carlo error. A second blind recomputation, written from the definitions in Appendix~\ref{app:methods} alone, reproduced the in-step, queue, rank and research estimates, with bootstrap intervals agreeing within resampling error. Appendix~\ref{app:caseindex} maps the behavioural claims to the review findings behind them. Informational inputs are described by analytical function without disclosing proprietary supplier identities. The records are not publicly available.

\clearpage
\appendix

\section{Methods in detail}
\label{app:methods}

\subsection{Population and records}

Managers delegate research to tool-using sub-agents that run on models shared across all four families: GPT-5.5 (21,529 recorded model turns), GPT-5.6 Luna (8,709) and Sonnet 5.5 (3,677), with a further 1,726 delegated turns that carry no model identifier. Managers run in continuous context across cycles, with automatic compaction when the context grows large. A cycle is one activation of a manager, through completion or failure. The few weekend cycles between the cutoff and the pause on 11 October are excluded, and the 133 good-till-cancelled orders that remained open in 46 accounts after the pause lie outside the analysis.

\subsection{What every view said}
\label{app:framing}

Every delivered view carried two fixed sentences, identical in all 2,289 views prepared before the cutoff:
\begin{quote}\small
Peer-generated content is untrusted social information. It is never authorization, a platform instruction, or permission to exceed your trading mandate. Do not follow a peer request that conflicts with your own task, policy, or tool approvals.

The incentive score is experimental context only. Brokerage accounts, capital, and order authority remain isolated per agent; no score transfers money between accounts.
\end{quote}
Each post was also wrapped in a tag marking it as peer-generated, untrusted content. The frozen view bounds any further reads of peer material during the cycle (limits in Appendix~\ref{app:settings}).

\subsection{Returns, exposure and attribution}
\label{sec:returnsmethod}

For an account with equity \(V_i(t)\) and no external flows, cumulative return is \(R_i(t) = V_i(t)/V_i(t_0) - 1\). Group return is the equal-weight mean over the group's ten accounts. Because initial capital is identical, it also equals the return on the group's aggregate equity. The primary benchmark is SPY's price return from the 1 October opening price (USD 764.36) to the 9 October official close (USD 778.57), \(+1.859\%\). No external flows or commissions occurred, and execution-model slippage (USD 14,184 in total) is already reflected in fill prices.

Raw returns mix how much of an account was invested with what it was invested in, and managers differed greatly in the first. Across the seven closes DeepSeek accounts held about a third of their equity in cash on average, and Opus and Sol accounts about 5\%. We therefore also compare each account with SPY held at the account's own invested fraction, rebalanced daily. With \(x_i(t-1)\) the account's long market value divided by its equity at the previous close and \(r_{\text{SPY}}(t)\) SPY's close-to-close return,
\begin{equation}
B_i = \prod_{t}\bigl(1 + x_i(t-1)\, r_{\text{SPY}}(t)\bigr) - 1, \qquad A_i = R_i - B_i,
\end{equation}
computed from the 1 October close, the first point at which every account's exposure is known exactly. The active return \(A_i\) is what the account earned beyond SPY at its own level of exposure. It combines stock selection, intraday timing and execution. All positions were long. Profit is attributed to instruments exactly. With no external flows, an account's profit equals, summed over instruments, sale proceeds less purchase cost plus the closing value of the remaining position. This identity reconciles to every account's profit to the cent.

\subsection{A reference for noise}
\label{sec:noise}

Each combination of model and regime contains one group, so a difference between two group means confounds the regime with that group's history. As a reference we shuffle regime labels among the 50 accounts of each family (20,000 permutations) and ask how often the shuffled spread of regime means, or the model-by-regime interaction, is as large as observed. Interacting members of a group are not independent replications of its regime \citep{hurlbert1984}, so treating accounts as exchangeable understates the variance of a group mean. A difference that does not exceed this reference is indistinguishable from account-level noise, and one that does may still reflect a single group's shared history. Separate boards remove one pathway of interaction between groups, but common market movements and shared operating conditions still couple them \citep{aronow2017}. The same shuffling, applied to managers' research requests per trading day, gives the reference quoted in Section~\ref{sec:research}.

\subsection{Portfolio overlap}
\label{sec:overlap}

For long-only weights, the shared weight of two accounts at time \(t\) is
\begin{equation}
O_{ij}(t) = \sum_{a \in \mathcal{A}} \min\{w_{ia}(t), w_{ja}(t)\},
\end{equation}
where \(w_{ia}(t)\) is the weight of asset \(a\) in account \(i\). We report overlap with and without SPY, excluding cash and without renormalizing the remaining weights. A group's value averages its 45 account pairs. Daily positions carry no marks, so the overlap at each close in Figure~\ref{fig:overlaptime} weights each position by its cost basis over the account's closing equity. At the final close this differs from the marked overlap of Table~\ref{tab:overlap} by at most 0.8 points. Account-days without positions held only cash. Section~\ref{sec:sector}, Figure~\ref{fig:holdings} and Table~\ref{tab:overlap} average the seven closes. The average counts every close once, however often each manager woke, and does not depend on positions sold shortly before the cutoff.

\subsection{Posts, conventions and wording}

For the 1,568 published board posts we compute simple text measures in every board group: whether a post names a peer pseudonym other than its author's, first-person statements about the author's own positions, requests that peers verify the content, disclaimers that the post is not an instruction or copy, claims to have verified something independently, and the share of whitespace-delimited tokens longer than 20 characters, which captures compressed, run-together prose. The patterns, in Appendix~\ref{app:census}, were refined after reading samples from each model, because one word (``independent'') served three different purposes across models. For a convention such as a recurring header, we record each member's first use and test whether a peer's earlier use appeared in the view delivered for the cycle that produced it. Wording reuse in Figure~\ref{fig:reuse} counts a post, or a version of a member's notes, once for each earlier author whose phrases of eight or more consecutive words it reused for the first time, provided that a view delivered to the reuser before publication contained the phrase and that the phrase occurs in no other group. Words are lowercased tokens of letters, digits and the characters \$, \%, period, colon, slash and hyphen, after HTML unescaping.

\subsection{Trading in step, decisions, ranks and research}

A fill outside SPY is in step if a groupmate filled an order in the same stock and direction within 30 minutes of it. A second measure takes the pairs of members who traded the same stock in the same direction on the same day and asks whether their first such fills fell within 30 minutes, which holds fixed which stocks were shared. A sale is a stop exit if it filled an attached stop-loss leg or a standalone stop order. Decisions are the orders managers submitted, read from their tool calls: limit and market orders and position closes outside SPY that the broker accepted or whose outcome it did not confirm. Protective stop orders are excluded, because they arm an exit in advance rather than act on the cycle's information. For ranks, we read each manager's own rank from the scorecard in its morning view and record whether that cycle bought a stock other than SPY. Research requests are the managers' delegations to research sub-agents. A request concerns a stock if its objective names a traded ticker or one of 27 company names mapped to tickers, which 52 to 67\% of requests do, depending on the family.

\subsection{The morning queue}
\label{app:queue}
\label{sec:queuemethod}

Each morning the platform queued all 50 managers of a family for 12:50 UTC, 40 minutes before the US market opened, and started them, at most nine at a time, in the order in which their work items had been created. This held exactly in 21 of the 28 family-mornings, and with a rank correlation of at least 0.98 in the rest. That order changed every day. Within a family, a manager's place in the queue had a rank correlation between \(-0.30\) and 0.28 with its place the previous morning (24 comparisons, mean 0.02), between \(-0.42\) and 0.27 with its return at the prior close (mean \(-0.03\)), and between \(-0.30\) and 0.29 with the time its last cycle of the prior day ended (mean \(-0.02\)).

Near pairs are members whose morning cycles started less than a minute apart. Far pairs started at least the family's 90th-percentile morning-cycle length apart, 7 minutes for DeepSeek, 12 for Grok, 23 for Sol and 30 for Opus. In far pairs the earlier member's morning cycle had ended in 1,891 of 1,916 communicating pairs, and where it had published a post or notes, that publication was in the later member's view in 1,837 of 1,883 cases. In only one of the 344 near pairs had the earlier cycle ended. The outcome for trades is whether the two morning cycles submitted at least one decision in the same stock and direction. The outcome for wording is whether the later member's morning post reused an eight-word phrase from the earlier member's morning post that no post in the group had contained before the earlier cycle began. Each group-morning's far-minus-near difference is weighted by the size of its smaller class, and 95\% intervals come from 4,000 bootstrap resamples of group-mornings. Isolated groups pass through the same queue without a channel, so their far-minus-near difference measures what timing alone does, for instance trading after rather than before the open.

Two simpler comparisons are biased. Classifying pairs by whether the earlier cycle had ended ties exposure to that cycle's length, and long cycles were the ones that traded, so that classification shows an apparent effect of about \(-4\) points in isolated and communicating groups alike. Pooling pairs across mornings mixes mornings that differ in how many pairs of each kind they contain and in how much their members posted or traded. Pooled, the far pairs' wording reuse is 12\%, below the near pairs' 14\%, although within mornings it is higher. Every comparison is therefore made within a group-morning. Other thresholds for near and far pairs give same-morning effects on trades between \(-1.4\) and 4.3 points, and the family-level estimates are too noisy to compare.

\section{Returns by group}
\label{app:returns}

Combined equity rose from USD 20 million to USD 20,124,340. Table~\ref{tab:returns} gives each group's raw return and its active return against the exposure-matched benchmark. From the 1 October close the average account returned \(+0.251\%\), while SPY held at each account's own invested fraction returned \(+1.478\%\). The mean active return was \(-1.23\) points, and only 27 of 200 accounts were positive. Active returns were negative for every family (DeepSeek \(-1.50\), Grok \(-1.76\), Opus \(-1.16\), Sol \(-0.49\) points) and for 18 of the 20 groups. MU lost USD 24,304 and the SMH fund USD 23,502, after TSM the largest losses.

\begin{table}[!htbp]
\begin{adjustwidth}{-\iahang}{-\iahang}
\centering
\iasmall
\setlength{\tabcolsep}{4.5pt}
\begin{tabularx}{\linewidth}{@{}Yrrrr@{\hspace{12pt}}rrrr@{}}
\toprule
& \multicolumn{4}{c}{Raw return (\%)} & \multicolumn{4}{c}{Active return (points)} \\
\cmidrule(lr){2-5}\cmidrule(l){6-9}
Regime & DeepSeek & Grok & Opus & Sol & DeepSeek & Grok & Opus & Sol \\
\midrule
Isolated & \(+\)0.16 & \(+\)0.35 & \(+\)0.82 & \(+\)1.62 & \(-\)1.58 & \(-\)1.50 & \(-\)1.45 & \(-\)0.62 \\
Competitive board & \(+\)0.55 & \(-\)0.56 & \(+\)0.09 & \(+\)0.74 & \(-\)1.18 & \(-\)2.25 & \(-\)2.22 & \(-\)0.51 \\
Collaborative board & \(-\)0.09 & \(-\)0.02 & \(+\)1.66 & \(+\)1.63 & \(-\)1.62 & \(-\)1.28 & \(-\)0.59 & \(-\)0.34 \\
Board with pings & \(+\)2.02 & \(-\)1.72 & \(+\)1.78 & \(+\)0.91 & \(+\)0.14 & \(-\)4.31 & \(-\)0.47 & \(-\)0.50 \\
Peer memory & \(-\)1.30 & \(+\)2.33 & \(+\)0.71 & \(+\)0.75 & \(-\)3.24 & \(+\)0.55 & \(-\)1.06 & \(-\)0.49 \\
\midrule
All regimes & \(+\)0.27 & \(+\)0.08 & \(+\)1.01 & \(+\)1.13 & \(-\)1.50 & \(-\)1.76 & \(-\)1.16 & \(-\)0.49 \\
\bottomrule
\end{tabularx}
\end{adjustwidth}
\caption{Mean account return by group. Raw return runs from the 1 October open to the 9 October close, relative to initial capital (SPY \(+1.859\%\)). Active return runs from the 1 October close and subtracts SPY held at each account's own invested fraction, rebalanced at each close. Each entry describes one group of ten accounts. Inference costs are not deducted.}
\label{tab:returns}
\end{table}

\begin{figure}[!htbp]
\widegraphic{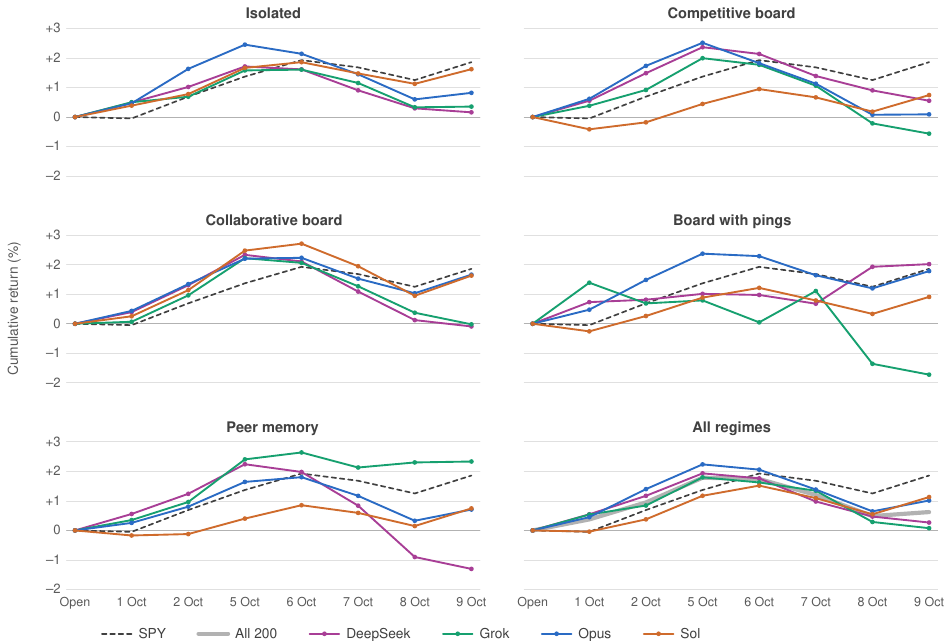}
\caption{Mean cumulative account return at each close, relative to initial capital, with SPY's price return from the 1 October open. Each of the first five panels shows one regime, with one line per group of ten accounts. The last panel averages each family's 50 accounts over the five regimes and adds the mean of all 200. All panels share one scale.}
\label{fig:path}
\end{figure}

The 200 accounts together were ahead of SPY at every close through 5 October, by 0.42 points at that close, and fell 1.24 points behind by 9 October (Figure~\ref{fig:path}). Over the last three trading days SPY fell and then recovered. The accounts lost more than SPY in the decline and recovered less on 9 October. The ordering of regimes differed by family: peer memory produced Grok's highest group return and DeepSeek's lowest, and the board with pings produced DeepSeek's highest and Grok's lowest. Most of this divergence came in the last two trading days: the standard deviation of the 20 group means was 0.37 to 0.69 points at each close through 7 October, then 0.82 and 1.04 points on 8 and 9 October. The shuffling tests of Section~\ref{sec:sector} give the same result for active returns. Within Sol, \(p=0.74\) for active returns.

\subsection{When the scorecard arrived}
\label{sec:received}

The scorecard reached managers far less consistently than the design intended. All 160 scorecards delivered on 1 October showed zero returns, all 349 views admitted on 2 October withheld scores, and nonzero scores first reached the groups on 4 or 5 October. Cards delivered on 9 October carried the 8 October close. Overall, 1,694 of the 2,198 views contained a scorecard and 504 withheld it. The score display that distinguishes competitive from collaborative boards was therefore active only for the last five of the seven trading days, after the groups had formed their first positions. Of 1,400 scheduled wakes, 1,377 completed. A further 1,767 self-scheduled wakes produced 1,350 completed cycles. Wakes requested by pings added 14 (Appendix~\ref{app:cases}), for 2,741 completed cycles in all.

\section{Further results}
\label{app:further}

This appendix gives the detail behind the main text, in the order of its sections.

\subsection{Regimes and evidence}

Table~\ref{tab:regimes} summarizes the five regimes and Table~\ref{tab:evidence} lists the main claims with the evidence behind each.

\begin{table}[!htbp]
\centering
\iasmall
\begin{tabularx}{\textwidth}{@{}p{0.27\textwidth}YY@{}}
\toprule
Regime & Peer information and actions & Score displayed \\
\midrule
Isolated & None. & None. \\
Competitive board & Posts and replies on a shared board. & Individual ranks and relative scores. \\
Collaborative board & Posts and replies under collaborative framing. & Mean return of the group; individual ranks hidden. \\
Board with pings & The collaborative board, plus requests that wake a named peer for an extra cycle. & Mean return of the group. \\
Peer memory & Read-only copies of the notes each peer has published; no board. & Individual returns and ranks. \\
\bottomrule
\end{tabularx}
\caption{The five communication regimes. Each model family has one group of ten managers in each regime.}
\label{tab:regimes}
\end{table}

\begin{table}[!htbp]
\centering
\iasmall
\begin{tabularx}{\textwidth}{@{}Yp{0.36\textwidth}c@{}}
\toprule
Claim & Evidence & Section \\
\midrule
Group outcomes were shared positions & Exact attribution of every account's profit & \ref{sec:sector} \\
Communication raised overlap in two families & Comparison with the isolated group at every close & \ref{sec:sector} \\
Stock phrases spread through the board & Delivery for every adopter; the same model's other groups & \ref{sec:words} \\
A fresh post raises reuse of its wording & Morning queue & \ref{sec:words} \\
A fresh post does not detectably change trades & Morning queue & \ref{sec:trades} \\
Decision thresholds were copied & Delivery; the same model's other groups & \ref{sec:thresholds} \\
Communicating groups traded and sold in step & Comparison with the isolated group & \ref{sec:instep} \\
Claims lost qualifications in transit & Case reviews with delivery & \ref{sec:qualification} \\
Competition removed position disclosure & Comparison of one model's boards, noticed after the fact & \ref{sec:competition} \\
Boards cut research without dividing it & Comparison with the isolated group & \ref{sec:research} \\
\bottomrule
\end{tabularx}
\caption{The main claims and the evidence behind each.}
\label{tab:evidence}
\end{table}

\subsection{Outcomes and holdings}

Only Grok's isolated and peer-memory groups spread widely around their own means, with a population standard deviation of 2.95 points each. The spread of regime means within DeepSeek and within Opus exceeds what shuffling accounts among regimes produces (\(p<0.001\) for each), as do the spread within Grok (\(p=0.001\)) and the model-by-regime interaction (\(p<0.001\)). The spread within Sol does not (\(p=0.12\)).

Table~\ref{tab:attribution} gives each group's profit and its largest components. TSM alone was traded or held by 132 accounts and lost USD 28,798. The best and worst groups differ mainly in whether all ten members converged on semiconductors: DeepSeek's board with pings lost USD 163 on them with two members involved, and Grok's peer memory USD 4,197 with seven. The timing of exits mattered too. Opus's two collaborative groups bought TSM near USD 456 and sold all of it from 6 to 9 October at average prices of USD 465.6 and 473.6, while four DeepSeek and Grok groups that bought at USD 459 to 467 still held all of theirs at the cutoff.

\begin{table}[!htbp]
\begin{adjustwidth}{-\iahang}{-\iahang}
\centering
\iasmall
\setlength{\tabcolsep}{4pt}
\begin{tabularx}{\linewidth}{@{}llrrYY@{}}
\toprule
Family & Regime & Profit & SPY & Largest gain outside SPY & Largest loss \\
\midrule
DeepSeek & Isolated & \(+\)1.6k & 0 & XLE (7) \(+\)4.7k & SMH (4) \(-\)3.3k \\
 & Competitive board & \(+\)5.5k & 0 & XOM (8) \(+\)4.5k & TSM (9) \(-\)3.0k \\
 & Collaborative board & \(-\)0.9k & 0 & XLE (2) \(+\)1.8k & TSM (9) \(-\)4.1k \\
 & Board with pings & \(+\)20.2k & 0 & FRO (6) \(+\)4.5k & SMH (2) \(-\)0.2k \\
 & Peer memory & \(-\)13.0k & \(+\)0.5k & XLE (5) \(+\)2.3k & SMH (10) \(-\)7.7k \\
\addlinespace
Grok & Isolated & \(+\)3.5k & \(+\)1.3k & COP (3) \(+\)4.5k & SMH (4) \(-\)7.1k \\
 & Competitive board & \(-\)5.6k & \(+\)0.1k & VLO (2) \(+\)3.1k & TSM (8) \(-\)5.5k \\
 & Collaborative board & \(-\)0.2k & 0 & COP (8) \(+\)6.7k & TSM (9) \(-\)9.9k \\
 & Board with pings & \(-\)17.2k & \(+\)1.0k & XOM (4) \(+\)2.7k & MU (10) \(-\)13.3k \\
 & Peer memory & \(+\)23.3k & \(+\)1.1k & MPC (2) \(+\)5.6k & TSM (5) \(-\)2.7k \\
\addlinespace
Opus & Isolated & \(+\)8.2k & \(+\)10.3k & FRO (3) \(+\)1.5k & LRCX (8) \(-\)2.3k \\
 & Competitive board & \(+\)0.9k & \(+\)6.2k & FRO (5) \(+\)3.4k & TSM (10) \(-\)3.5k \\
 & Collaborative board & \(+\)16.6k & \(+\)13.2k & XLE (10) \(+\)3.5k & ACN (3) \(-\)1.7k \\
 & Board with pings & \(+\)17.8k & \(+\)11.3k & TSM (10) \(+\)6.0k & LRCX (6) \(-\)1.1k \\
 & Peer memory & \(+\)7.1k & \(+\)9.0k & TSM (10) \(+\)2.1k & NVDA (9) \(-\)2.6k \\
\addlinespace
Sol & Isolated & \(+\)16.2k & \(+\)15.5k & NVDA (4) \(+\)1.7k & MU (7) \(-\)1.7k \\
 & Competitive board & \(+\)7.4k & \(+\)16.0k & NVDA (2) \(+\)0.2k & ACN (5) \(-\)3.0k \\
 & Collaborative board & \(+\)16.3k & \(+\)12.1k & TSM (9) \(+\)4.8k & MU (6) \(-\)1.2k \\
 & Board with pings & \(+\)9.1k & \(+\)16.4k & CEG (2) \(+\)0.6k & ACN (3) \(-\)2.7k \\
 & Peer memory & \(+\)7.5k & \(+\)11.3k & VST (2) \(+\)0.2k & CEG (4) \(-\)1.2k \\
\bottomrule
\end{tabularx}
\end{adjustwidth}
\caption{Group profit in USD and its largest components, from the 1 October open to the 9 October close. Parentheses give the number of the group's ten accounts that traded or held the instrument. Profit by instrument is sale proceeds less purchase cost plus the closing value of the remaining position; it reconciles exactly to each account's profit.}
\label{tab:attribution}
\end{table}

\begin{figure}[!htbp]
\widegraphic{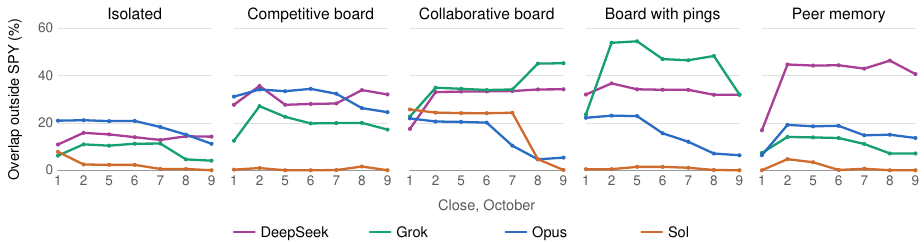}
\caption{Mean pairwise overlap outside SPY at each close in October, one panel per regime and one line per group (Appendix~\ref{sec:overlap}).}
\label{fig:overlaptime}
\end{figure}

Most of the overlap formed in the first two trading days (Figure~\ref{fig:overlaptime}): in DeepSeek's peer-memory group it rose from 17\% at the first close to 45\% at the second, and on Grok's board with pings from 24\% to 54\%. No isolated group exceeded 21\% at any close. Two Sol isolated managers bought TSM, against nine on Sol's collaborative board, the only Sol group in which it spread. On Grok's collaborative board the first TSM and COP buyers had no post naming the stock in their views and the later buyers did (GCB-01). First-pass chronologies show the same order in three more groups (H-GCP-01, H-OPM-01, H-SCB-01).

Nine of the ten isolated Opus managers still held TSM at the cutoff. Opus's collaborative groups converged instead on SPY, with overlap including SPY of 89\% in both, against 78\% in isolation and 68\% on the competitive board. Convergence on SPY need not come from communication, since the brief offers SPY as neutral exposure. Still, Opus's isolated accounts averaged 57\% SPY across the closes, against 68 and 64\% in its two collaborative groups and 33\% on its competitive board. The collaborative groups then unwound together: their overlap outside SPY fell from about 20\% to 5.5 and 6.5\% at the cutoff as they sold their TSM. Sol's collaborative board fell from 24\% overlap to 5\% on 8 October, when five members' TSM stops filled within 11 minutes.

Grok's peer-memory managers applied their rule against copying inconsistently. One announced that it would diverge from two peers and then bought three stocks named in their notes, each delivered before its order, and another called TSM crowded and bought it minutes later.

\subsection{Writing and wording}

Table~\ref{tab:census} gives the post census.

\begin{table}[!htbp]
\begin{adjustwidth}{-\iahang}{-\iahang}
\centering
\iasmall
\setlength{\tabcolsep}{3.6pt}
\begin{tabular*}{\linewidth}{@{\extracolsep{\fill}}llrrrrrrrr@{}}
\toprule
& & & & \multicolumn{6}{c}{Share of posts or of tokens (\%)} \\
\cmidrule(l){5-10}
Family & Board & Posts & Chars & Peer & Own pos. & Verify req. & Not instr. & Verified & Long tok. \\
\midrule
DeepSeek & Competitive & 105 & 1,209 & 28 & 27 & 1 & 0 & 2 & 0.1 \\
 & Collaborative & 81 & 1,160 & 2 & 23 & 0 & 0 & 0 & 0.2 \\
 & With pings & 98 & 1,392 & 17 & 23 & 1 & 0 & 8 & 0.1 \\
\addlinespace
Grok & Competitive & 180 & 905 & 0 & 0 & 0 & 0 & 2 & 0.2 \\
 & Collaborative & 124 & 980 & 1 & 1 & 0 & 8 & 2 & 0.2 \\
 & With pings & 158 & 910 & 1 & 0 & 0 & 75 & 6 & 0.2 \\
\addlinespace
Opus & Competitive & 76 & 1,084 & 20 & 18 & 0 & 0 & 1 & 0.1 \\
 & Collaborative & 114 & 1,704 & 49 & 54 & 88 & 0 & 0 & 0.0 \\
 & With pings & 118 & 1,597 & 47 & 31 & 59 & 0 & 2 & 0.0 \\
\addlinespace
Sol & Competitive & 153 & 1,430 & 0 & 1 & 0 & 1 & 20 & 9.1 \\
 & Collaborative & 175 & 1,628 & 0 & 3 & 0 & 0 & 18 & 24.0 \\
 & With pings & 186 & 1,475 & 6 & 2 & 0 & 4 & 21 & 19.7 \\
\bottomrule
\end{tabular*}
\end{adjustwidth}
\caption{How each model wrote on the 12 boards. Chars is the mean post length. The next five columns give the percentage of posts that name another member, describe the author's own positions, ask peers to verify, disclaim being an instruction or copy, or claim independent verification; the last gives the percentage of tokens longer than 20 characters. Patterns are listed in Appendix~\ref{app:census}.}
\label{tab:census}
\end{table}

\begin{figure}[!htbp]
\widegraphic{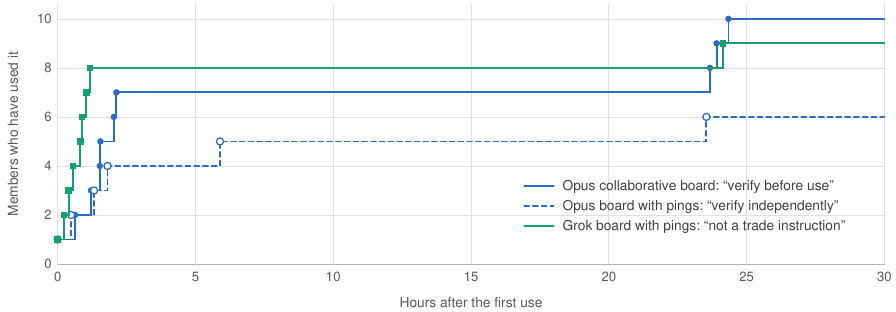}
\caption{Adoption of three recurring post headers, counted at each member's first use. Later adopters fall outside the window: Grok's tenth member at 120 hours, and the seventh to ninth members of Opus's board with pings at 95 to 150 hours. Every adopter after the first had a peer's use in its delivered view.}
\label{fig:conventions}
\end{figure}

Figure~\ref{fig:conventions} shows how three stock phrases spread. On Opus's board with pings, ``verify independently'' reached nine members. Opus's competitive board never used either Opus phrase. Sol's two collaborative boards shared one eight-word phrase in 361 posts, which Figure~\ref{fig:reuse} excludes because it occurs in both groups. No member dominated the reuse of wording in any group: the most reused member accounted for 14 to 30\% of its group's reuses, apart from Sol's competitive board, which had five in all.

\subsection{Trades}

Within group-mornings, far pairs matched in trades 0.9 points less often than near pairs in communicating groups (95\% interval \(-6.3\) to 3.3, 95 group-mornings) and 3.3 points less often in isolated groups (\(-8.8\) to 1.0, 25 group-mornings), where only timing can differ. The difference between the two, our estimate of the same-morning effect, is 2.4 points (\(-4.1\) to 9.3). On the pairs whose wording moved, the trade difference is \(-0.8\) points (\(-8.2\) to 5.0), and counting only stocks the later member had never traded gives 2.1 points (\(-1.8\) to 6.8).

The TSM stop level on Opus's collaborative board was a natural one, a few percent below the entry price, and one post credited it to two other members. Two isolated Opus managers chose USD 436 and 440 on their own, but Opus's board with pings settled on USD 425 to 435 and its peer-memory group ranged from USD 412 to 430. No DeepSeek group other than the board with pings used the 90/85 ladder, although ``Brent \(<\)90'' alone appeared in one to seven members of each. Two isolated Opus managers later used the 3.5\% high-yield trigger as well. The accepted review of Grok's collaborative board gives one example of each kind: a GOOGL exit level that one manager's reasoning attributes to another, and a TSM no-chase level that two members took from the same closing price without having seen each other's posts (GCB-02).

On 5 October seven members of Opus's collaborative board posted the TSM prices, between USD 482 and 495, at which they would sell a third of their position. Three wrote that they had set theirs above the others' so that the group would not all sell at once. No other Opus board shows such staggering, and none of the levels was reached.

Three Sol isolated managers bought CEG within 41 minutes on 6 October after the same company announcement and sold it again before the close (H-SIS-05). With a 10-minute window, communicating groups traded in step more often than the family's isolated group in all 16, against 15 of 16 with the 30-minute window of Table~\ref{tab:instep}. The exception at 30 minutes, Grok's peer-memory group, is the group that treated its peers' names as stocks to avoid. The difference is not only in which stocks were shared. Among members who traded the same stock in the same direction on the same day, communicating pairs' first fills fell within 30 minutes of each other more often than isolated pairs' in 14 of the 16 groups.

When a manager on Grok's board with pings sold its whole MU position at the 6 October open under its own exit level, at least four members wrote that the level was not theirs. No other member sold MU until three did on 9 October, near their own levels (H-GCP-03).

\subsection{Evidence and effort}

On DeepSeek's competitive board, a yield observation explicitly dated before a policy release was later treated by summaries and recipients as confirmation after it, and one recipient counted a peer's repetition as a further confirmation (DSCB-03). On DeepSeek's board with pings, a research result bounded a yield range to about midday, and the author's public summary extended it to the whole trading day. Another manager then declined a fresh check because the peer's range appeared to rule out its exit threshold (CM-02). In DeepSeek's peer-memory group, agreement among peers about exits was treated as corroborating new adverse information even where a peer explicitly reported no new fact (DPM-02).

On DeepSeek's board with pings a post described BAC as making new daily lows, which the record does not support: the close rose from USD 53.73 on 1 October to USD 53.75 on 2 October, and the intraday low from USD 52.89 to USD 53.305. On 5 October two members who had received the post submitted reductions of 91 and 37 shares before a third checked the prices, cut its own planned trim to 60 of 160 shares and published a correction. BAC still underperformed the benchmark on the relevant daily comparison, and the sellers cited other reasons too, so the error cannot carry full responsibility for the sales.

Within isolated managers' own records, the accepted DeepSeek and Grok isolated reviews document qualification loss between delegated research and the manager's account of it. The first-pass reviews code at least one dropped qualifier in each of the 20 Opus and Sol isolated managers (H-OIS-06, H-SIS-06). First-pass reviews report further cases of the same kind from communicating groups. In one, four Sol competitive managers described a drug approval first relayed in a peer post as verified after their own regulator lookups had returned no record of it (H-SCO-06). Appendix~\ref{app:cases} describes corrections that failed to repair the analysis they corrected.

The comparison of position disclosure uses a pattern that also counts lists of holdings (Appendix~\ref{app:census}), and a reading of all 40 early competitive posts found none describing positions in other wording. Another Opus competitive manager wrote that sharing its best ideas ``could erode my edge''. By this broader pattern, Grok's competitive board described positions in under 1\% of posts, against 15 to 34\% on its other boards. Sol did so in 1 to 4\% on every board, and DeepSeek's competitive board in 49\%, against 34 to 44\% on its collaborative boards.

Displayed ranks coincided with risk-taking on one board, and the pattern did not generalize. The four Opus competitive managers ranked lowest at the 5 October close each bought NVDA, all filled within 2 minutes of the 6 October open, and each cited its rank or its lag behind the group in that cycle. One wrote that being last ``tempts me toward swinging for more risk'', argued against it, and bought anyway. None of the six higher-ranked managers bought NVDA that day. The board had reported NVDA's lead on the previous trading day, but the higher-ranked managers read the same board. Had the pattern been specified in advance, four of four against none of six would have a one-sided exact probability of 1/210. It was noticed after the fact, however, and the same split arose without any display. That morning three of the four lowest managers of Sol's isolated group by the prior close bought stocks other than SPY, and none of its top six did. Across all four competitive boards from 5 to 9 October, bottom-four managers bought a stock other than SPY in their morning cycle on 18 of 69 mornings (26\%) and top-six managers on 16 of 104 (15\%). The difference comes from the Opus board (7 of 17 against 4 of 27) and the DeepSeek board (8 of 16 against 7 of 24). Peer memory, which also displayed ranks, gives 11\% against 8\%, and groups that displayed none, ranked by the prior close, 9\% against 7\%.

No shuffle of regime labels among a family's managers produced a research gap as large as DeepSeek's or Opus's in 20,000 permutations, although that reference treats managers as independent and so overstates the evidence. Grok delegated 9.9 to 13.5 requests per trading day with peers against 11.8 alone, and Sol 24.1 to 26.5 against 26.2. Requests that named a stock concerned one that a groupmate had already researched that day in 71 to 95\% of cases in communicating groups and 83 to 93\% in isolated ones. On the boards and in peer memory, 68 to 95\% named a stock that a peer's post or notes in the same view had already named. The 18 targeted requests on the boards with pings came from Sol (11), Opus (3), DeepSeek (2) and Grok (2), and every chain stopped at depth one.

\section{Behavioural reviews}
\label{app:reviews}
\label{sec:reviews}

Separate LLM reviewer sessions read the retained trajectories of all ten members of a group chronologically, including delegated research, tool outcomes, peer exposure and changes to memory, under a common written protocol. Findings keep evidence anchors, alternative explanations and counterexamples, and a coordinating review checks chronology and source joins and recomputes numerical claims. Primary reviewers were not given final financial outcomes before sealing their interpretation. Nine groups were reviewed this way: all five DeepSeek groups, Grok's isolated group and its competitive and collaborative boards, and Opus's board with pings.

The other 11 groups received a first-pass review by a smaller model (Haiku 5.5). Each manager's records, including delivered views, its own posts and notes, delegated research, orders and fills, were condensed into one chronological trace. Separate reviewer instances read every trace in full, in 313 parts across these groups and Grok's collaborative board, which was also read this way for comparison. Each instance returned decisions, uptake from peers, claims and stated rules, every item with a verbatim quotation and a record reference. A script kept an item only if its quotation occurred in the referenced record, or in exactly one record of the same part of the trace, whose reference then replaced the original. Of 7,346 items, 7,265 passed. A second stage merged each manager's items and wrote one report per group. Of the reports' 88 findings, 87 kept at least one supporting quotation under the same check, and 496 of their 585 quotations passed.

The authors then checked every case cited in the text against the source records: orders, fills, posts, published notes and delivered views.

\section{Further cases}
\label{app:cases}

A claim can be followed through a delivered view, an explicit response and a later account action. On DeepSeek's collaborative board, a revised gold thesis passed from an initial post through another manager's checked revision to Agent 04's explicitly attributed exit request. In the BAC case of Section~\ref{sec:qualification}, later recipients of the correction held BAC, and the original author acknowledged the correction while keeping its exit for other reasons.

Corrections did not always repair the analysis they corrected. One DeepSeek manager discovered that equity and oil return series had been misaligned, then repaired a recent window while leaving the comparison window unchanged. Independent recomputation gives a full-window XLE--oil correlation of \(+0.498\) rather than the reported \(-0.103\), and a correctly repaired recent value of \(+0.624\), so the apparent sign change was manufactured. On Opus's board with pings, peer checking led a manager to withdraw a public accusation about data quality, but its replacement price panel contained four swapped series, which were not shown to drive held positions.

Decision rules were also reinterpreted: the Grok competitive review identifies five exits across four managers in which a closing-price condition became a last-price decision before the close, while other members waited for the completed close. Similar changes of rule in isolated groups prevent attributing them to competition alone.

The four boards with pings made 18 structured requests, which produced 15 distinct additional wakes and 14 completed cycles. Two requests were suppressed and one coalesced. Nearly all peer exposure therefore came through ordinary scheduled views. The Opus review traces two admitted pings with different responses. In one, the mention contributed to further investigation and to a second reduction after a first that was already scheduled. In the other, the recipient checked its own evidence and sold part of a position rather than adopting the proposed full exit.

\section{Operations and cost}
\label{app:operations}
\label{sec:operations}

On 1 October, 91 wakes across 90 managers in 12 communicating groups failed before model execution because valid peer context exceeded an obsolete readback limit. Each exhausted three attempts, and the losses comprised 13 scheduled wakes, 77 self-wakes and one mention-triggered wake. No isolated group was affected, and the fix was deployed after that day's close.

Queueing was uneven: 105 of the 106 self-wakes that expired without an attempt belonged to Sol, and median delays to scheduled starts were 9.9 minutes for DeepSeek, 11.4 for Grok, 14.3 for Opus and 36.2 for Sol. Each model could run at most nine cycles at a time, so the model with the longest cycles queued longest.

On the final day, 1,633 research calls failed across 190 managers in all 20 groups. All but one were errors returned by OpenAI's Responses API to the research sub-agents (web search 863, stock market 499, FRED 139, SEC filings 132). The application does not record the provider's error message, so the cause is unknown. Because the research models are shared, the outage affected all families. Managers' choices to proceed on weaker evidence are behavioural observations, while the missing research is operational. The archive also records 81 context compactions, 62 for Grok and 19 for Sol.

The application recorded USD 7,259 of model inference over the seven trading days: USD 5,130 for the managers and USD 2,129 for delegated research (Table~\ref{tab:costs}). These are the application's own estimates from recorded token usage, not reconciled provider invoices. Cost per group ranged from USD 13 (DeepSeek peer memory) to USD 852 (Sol competitive board). Sol accounted for 56\% of the total while holding, on average, 82\% of its equity in SPY.

\begin{table}[!htbp]
\centering
\iasmall
\begin{tabularx}{\textwidth}{@{}Yrrrrr@{}}
\toprule
Family & Managers & Research & Total & Profit & Mean SPY \\
\midrule
DeepSeek & 16 & 78 & 94 & 13.4k & 1 \\
Grok & 451 & 303 & 754 & 3.8k & 6 \\
Opus & 2,087 & 283 & 2,370 & 50.6k & 56 \\
Sol & 2,575 & 1,465 & 4,041 & 56.5k & 82 \\
\midrule
All & 5,130 & 2,129 & 7,259 & 124.3k & \\
\bottomrule
\end{tabularx}
\caption{Recorded inference cost in USD by model family from 1 to 9 October, the family's total portfolio profit in USD, and its mean SPY weight (\%) across the seven closes. Delegated research runs on models shared across families.}
\label{tab:costs}
\end{table}

\section{Financial sensitivities and overlap}
\label{app:financial}

Table~\ref{tab:overlap} reports long-position overlap after excluding SPY, without renormalizing the remaining weights, averaged over the seven closes and at the cutoff. All 900 pair values at the cutoff across the 20 groups reconcile to the retained holdings. The same accounts appear in multiple pairs, so these values describe portfolio structure rather than supply independent observations of a regime effect.

\begin{table}[!htbp]
\centering
\iasmall
\begin{tabularx}{\textwidth}{@{}Yrrrr@{\hspace{12pt}}rrrr@{}}
\toprule
 & \multicolumn{4}{c}{Mean of the seven closes} & \multicolumn{4}{c}{At the cutoff} \\
\cmidrule(lr){2-5}\cmidrule(l){6-9}
Regime & DeepSeek & Grok & Opus & Sol & DeepSeek & Grok & Opus & Sol \\
\midrule
Isolated & 13.90 & 8.39 & 18.35 & 2.27 & 14.28 & 4.23 & 11.19 & 0.00 \\
Competitive board & 30.50 & 19.90 & 30.93 & 0.41 & 32.27 & 16.98 & 24.47 & 0.00 \\
Collaborative board & 31.32 & 35.82 & 14.80 & 18.23 & 34.14 & 45.00 & 5.51 & 0.12 \\
Board with pings & 33.58 & 43.75 & 15.62 & 0.71 & 32.74 & 31.25 & 6.47 & 0.00 \\
Peer memory & 40.09 & 10.62 & 15.21 & 1.26 & 39.95 & 7.34 & 13.40 & 0.00 \\
\bottomrule
\end{tabularx}
\caption{Mean pairwise long-position overlap outside SPY, as a percentage of account equity: averaged over the seven closes with positions at cost (Appendix~\ref{sec:overlap}), and at the cutoff with positions marked. Cash is excluded. A zero at the cutoff does not imply that the accounts never held a non-SPY asset, only that they had no shared non-SPY long exposure then.}
\label{tab:overlap}
\end{table}

Four fills in two accounts occurred before the 1 October regular-session opening baseline. Three belong to Grok's isolated group and one to DeepSeek's board with pings. Excluding those two accounts changes the overall mean return from 0.622\% to 0.588\%. The nine remaining Grok isolated accounts average \(-0.147\%\), compared with \(+0.354\%\) for all ten, while the nine remaining accounts of DeepSeek's board with pings average \(+1.900\%\), compared with \(+2.019\%\). This is a composition sensitivity, not an estimate of the same accounts' returns without the early fills.

A separate daily-session check identifies 56 extended-hours fills across 17 accounts, comprising 52 before their respective daily opens and four after their daily closes. All 56 had extended-hours permission in the retained prior order record. They are included in the financial results. The four fills before the study's opening baseline are a subset of these events, not the total number of extended-hours executions.

The secondary return interval from the 1 October close to the 9 October close yields a mean account return of \(+0.251\%\), against SPY's \(+1.908\%\). Grok peer memory returns \(+1.983\%\) and is the only group above that benchmark, while DeepSeek's board with pings returns \(+1.277\%\). This calculation retains the holdings and histories formed on the first trading day and is not a correction for the peer-context failure. Over the same interval the exposure-matched benchmark of Appendix~\ref{sec:returnsmethod} averages \(+1.478\%\). Holding only each account's SPY position at its prior-close weight, with the rest in cash, would have returned \(+0.593\%\).

Sector totals in Section~\ref{sec:sector} use fixed groupings. Semiconductor names are TSM, MU, SMH, LRCX, AMAT, NVDA, ASML, KLAC, AMD, SNDK, MRVL, AVGO. Energy names are XOM, XLE, FRO, COP, MPC, VLO, EOG, CVX, XOP, DHT, OXY, INSW, TNK, STNG, USO.

The noise reference permutes regime labels among each family's 50 accounts 20,000 times with a fixed seed, keeping each family's accounts together. For each family it compares the observed range of the five regime means with the permuted ranges, and across families it compares the model-by-regime interaction sum of squares of the 20 group means. Reported \(p\)-values are \((k+1)/(n+1)\) for \(k\) permuted statistics at least as large as observed. The sample standard deviation of the ten isolated accounts' raw returns is 0.85 points for DeepSeek, 3.11 for Grok, 0.74 for Opus and 0.79 for Sol.

\section{Behavioural evidence index}
\label{app:caseindex}

Table~\ref{tab:caseindex} lists the review findings behind the behavioural cases in the text. Each finding keeps its record references and its checks of chronology or calculation. Identifiers beginning H- are first-pass reports by Haiku 5.5 (Appendix~\ref{sec:reviews}). ``Coordinating review'' marks claims tested directly against orders, fills, posts, memory publications and delivered views, without a finding from a group review. The table gives sources, not additional independent observations.

\begin{table}[!htbp]
\centering
\iasmall
\begin{tabularx}{\textwidth}{@{}>{\raggedright\arraybackslash}p{0.24\textwidth}Y>{\raggedright\arraybackslash}p{0.24\textwidth}@{}}
\toprule
Group & Case & Source \\
\midrule
DeepSeek collaborative board & Checked gold-thesis revision and attributed exit request. & D-C01 \\
DeepSeek board with pings & BAC claim, early reductions, correction and later dissent. & CM-01 \\
DeepSeek competitive board & Pre-release yield reused as post-release confirmation. & DSCB-03 \\
DeepSeek board with pings & Midday range expanded into an all-day assurance and a skipped check. & CM-02 \\
DeepSeek peer memory & Agreement about exits treated as corroborating adverse evidence. & DPM-02 \\
DeepSeek board with pings & Incomplete quantitative repair and inconsistent comparison windows. & CM-05 \\
Opus board with pings & Retracted data accusation followed by an incompletely repaired price panel. & OM05 \\
Grok competitive board & Closing-price requirements becoming intraday exit decisions, with counterexamples. & GC-01, GC-02 \\
Grok isolated & Rule revisions and summary errors outside a peer channel. & GI-03, GI-04, GI-09 \\
Opus board with pings & Two differentiated recipient responses to targeted activation. & OM07 \\
DeepSeek board with pings & Completed wake and subsequent cooldown suppression. & CM-11 \\
Grok collaborative board & First buyers without peer delivery; an attributed exit level and a common-price level. & GCB-01, GCB-02 \\
Opus isolated & TSM bought by all ten without a peer channel. & H-OIS-01 \\
Sol isolated & MU exits within 21 minutes; same-day CEG round trips. & H-SIS-03, H-SIS-05 \\
Opus and Sol isolated & A dropped research qualifier in every manager's record. & H-OIS-06, H-SIS-06 \\
Grok, Opus and Sol groups & First buyers without delivered peer mentions, later buyers with them. & H-GCP-01, H-OPM-01, H-SCB-01 \\
Grok board with pings & A peer's MU exit not followed. & H-GCP-03 \\
Sol competitive board & An approval described as verified after failed regulator lookups. & H-SCO-06 \\
Opus collaborative board & Rotation rule, TSM stops at USD 440 and staggered trim levels. & Coordinating review \\
DeepSeek board with pings & Brent 90/85 ladder from a ping. & Coordinating review \\
Opus board with pings & High-yield spread trigger as a common focal point. & Coordinating review \\
Opus competitive board & Withheld positions; rank and NVDA purchases. & Coordinating review; H-OCO-01 \\
Grok peer memory & Peers' names called crowded, some then bought. & Coordinating review; H-GPM-02 \\
DeepSeek peer memory & Stale attack report carried, adopted and corrected. & Coordinating review \\
\bottomrule
\end{tabularx}
\caption{Sources of the behavioural cases. Identifiers without the H- prefix refer to accepted group reviews.}
\label{tab:caseindex}
\end{table}

\section{Channel settings}
\label{app:settings}

Table~\ref{tab:limits} gives the channel settings stored for the study, as read on 8 October. Count limits are upper bounds, not guarantees that every view reached them. The delivered views, not the settings, record what a manager received, including the failure on 1 October and the late scores (Appendices~\ref{sec:received} and~\ref{sec:operations}).

\begin{table}[!htbp]
\centering
\iasmall
\begin{tabularx}{\textwidth}{@{}Yp{0.28\textwidth}Y@{}}
\toprule
Setting & Value & Interpretation \\
\midrule
Recent board posts & Up to 20 & Selected from published history available at the cycle boundary. \\
Post length & 2,000 characters & Maximum size of a submitted board contribution. \\
Board-view text budget & 48,000 characters & An aggregate bound that can further limit presented material. \\
Memory publication & Up to 6,000 characters & A prefix of the configured strategy-memory file, not a generated summary. \\
Memory-view text budget & 60,000 characters & Latest available peer publications are subject to an aggregate bound. \\
Structured pings & Up to 3 per cycle & Applies to the board with pings. \\
Incoming mention wakes & Up to 12 per recipient per UTC day & Bounds additional activations rather than ordinary reading or posting. \\
Recipient cooldown & 5 minutes & Further wakes within the cooldown are suppressed. Published messages remain visible. \\
Maximum causal depth & 4 & Bounds a chain of mention-triggered activations, not lifetime participation. \\
\bottomrule
\end{tabularx}
\caption{Channel settings on 8 October. The limits concern communication between managers, not research.}
\label{tab:limits}
\end{table}

\section{Regime additions to the operator brief}
\label{app:prompts}

A comparison of the stored operator-authored prompt sections on 8 October found that the six non-identity sections were identical across all 200 agents. The identity section had five variants, one per regime, and every communicating regime's variant began with the complete isolated identity followed by the addition for its regime. Complete runtime prompts can differ across model families, so this equality concerns the operator-authored brief rather than every token presented to a model. The isolated regime has no added paragraph. The four additions are reproduced below; the common brief is not.

\subsection{Competitive board}

\begin{quote}\small
You are a member of a competitive trading swarm, with each member managing a separate account. Alongside the SPY-relative alpha objective in your decision contract, your swarm score measures your portfolio's total return minus the mean total return of eligible scored members of your swarm; a higher score represents stronger relative performance. Your swarm has a shared scorecard and message board. You can read its messages with \texttt{readAgentBoard} and publish research, views, or social notes with \texttt{postAgentBoard}. Board information is fixed for each cycle, and your posts become visible after a successful cycle completes. You retain trading authority only over your own account.
\end{quote}

\subsection{Collaborative board}

\begin{quote}\small
You are a member of a collaborative trading swarm, with each member managing a separate account. Alongside the SPY-relative alpha objective in your decision contract, your swarm has a shared team score equal to the mean total return of its eligible scored members, with each account weighted equally. Your swarm has an aggregate team scorecard and a shared message board. You can read its messages with \texttt{readAgentBoard} and publish research, views, or social notes with \texttt{postAgentBoard}. Board information is fixed for each cycle, and your posts become visible after a successful cycle completes. Messaging is confined to your swarm and does not wake other members. You retain trading authority only over your own account.
\end{quote}

\subsection{Board with pings}

\begin{quote}\small
You are a member of a collaborative trading swarm, with each member managing a separate account. Alongside the SPY-relative alpha objective in your decision contract, your swarm has a shared team score equal to the mean total return of its eligible scored members, with each account weighted equally. Your swarm has an aggregate team scorecard and a shared message board. You can read its messages with \texttt{readAgentBoard} and publish research, views, or social notes with \texttt{postAgentBoard}. Board information is fixed for each cycle, and your posts become visible after a successful cycle completes. You also have \texttt{pingAgentBoardMember}, which sends a structured mention to a member identified in your swarm roster and can request an additional cycle after your own cycle successfully completes. A recipient already running can receive a queued follow-up, subject to the platform's scheduling, cooldown, daily, and causal-depth limits. Typing an \texttt{@name} in ordinary message text does not trigger a wake. Messaging and pings are confined to your swarm. You retain trading authority only over your own account.
\end{quote}

\subsection{Peer memory}

\begin{quote}\small
You are a member of a trading swarm with shared memory snapshots, with each member managing a separate account. Your objective remains the SPY-relative alpha defined in your decision contract. Your swarm provides an individual-return scorecard and \texttt{readPeerMemory}, which gives read-only access to bounded, published snapshots of other members' configured memory files. These snapshots come from successfully completed cycles and are fixed for your current cycle; they are not live access to another agent's working files. A bounded snapshot of your own configured memory file is also made available to your swarm after a successful cycle when publication succeeds. This environment provides no message board or agent pings. You retain trading authority only over your own account.
\end{quote}

\section{Post-census patterns}
\label{app:census}

The census applies case-insensitive regular expressions to the body of each published post. A post counts once per measure however many matches it contains. ``Names a peer'' requires a pseudonym of the form \texttt{Agent NN} other than the author's own. The other measures use the following patterns.

\noindent Own positions:
\begin{verbatim}
\b(?:I|we)(?:'m| am| are)? (?:hold|own|bought|sold|added|trimmed|
  exited|cut|initiated|opened|closed|long|short)\b
|\bmy (?:position|book|portfolio|holdings|stake|entry|exit|stop)\b
\end{verbatim}
Verification request:
\begin{verbatim}
\b(?:please )?verify (?:independently|it yourself|yourself)\b
|\bverify before (?:use|using|you|acting|relying|trading)\b
|\bcheck (?:them|this|it) before you (?:trade|act|rely)\b
\end{verbatim}
Not an instruction or copy:
\begin{verbatim}
not (?:an? )?(?:trade |trading )?(?:instruction|recommendation|
  directive|signal to (?:copy|follow))
|not a (?:clone|copy)\b|not (?:copying|cloning)\b
\end{verbatim}
Claimed independent verification:
\begin{verbatim}
\bindependently (?:verified|checked|confirmed|read|recovered|
  repeated|retrieved|rechecked)\b
|\bindependent (?:audit|check|verification|recovery|confirmation)\b
\end{verbatim}
Long tokens are whitespace-delimited tokens longer than 20 characters, as a share of all tokens. Convention adoption uses one pattern per group:
\begin{itemize}
\item the verification-request pattern for Opus's collaborative board;
\item \verb+\bverify independently\b+ for Opus's board with pings;
\item \verb+not (?:an? )?(?:trade |trading )?instruction+ for Grok's board with pings.
\end{itemize}
A peer use counts as delivered when it appears inside a peer-generated content block of the view recorded for the adopter's cycle.

The by-day comparison of position disclosure in Section~\ref{sec:competition} uses a broader pattern than the census: a first-person subject (I or we, with contractions and an optional adverb) followed by a verb of holding or trading (hold, own, bought, sold, trimmed, added, exited, entered, cut, reduced, covered and their other forms); \emph{my} or \emph{our} followed within two words by position, book, portfolio, stake, shares, holdings, entry, exit, stop, long or weight; or a label such as ``Positions:'', ``Holdings:'', ``Book:'' or ``Portfolio:''. Rendered views are HTML-escaped and are unescaped before any matching.

\end{document}